\documentclass[12pt]{article} 

\usepackage{amsmath,amssymb,amsfonts}
\usepackage{psfrag}
\usepackage{color}
\definecolor{darkblue}{rgb}{0.1,0.1,.7}
\usepackage[colorlinks, linkcolor=darkblue, citecolor=darkblue, urlcolor=darkblue, linktocpage]{hyperref} 
\usepackage[square, comma, compress,numbers]{natbib}
\usepackage[]{amsmath}
\usepackage[]{graphicx}
\usepackage[]{latexsym}
\usepackage{geometry}
\usepackage{amscd}
\usepackage[all,cmtip]{xy}
\usepackage{mathrsfs}
\usepackage[margin=10pt,font=small,labelfont=bf]{caption}
\usepackage{dsdshorthand}
\usepackage{simplewick}
\usepackage{changepage}
\numberwithin{equation}{section}

\renewcommand{\be}{\begin{eqnarray}}
\renewcommand{\ee}{\end{eqnarray}}
\newcommand{\bea}{\begin{eqnarray}}
\newcommand{\eea}{\end{eqnarray}}

\def\beq{\begin{equation}} 
\def\eeq{\end{equation}} 
\def\<{\langle}
\def\>{\rangle}

\def\nn{\nonumber} 
 
\def\cO {{\cal O}} 
\def\cG {{\cal G}} 
\def\cF {{\cal F}} 
\def\cN {{\cal N}}

\begin{document}

\vspace*{-.6in} \thispagestyle{empty}
\begin{flushright}
\end{flushright}
\vspace{.2in} {\Large
\begin{center}
{\bf Bootstrapping Mixed Correlators \\ in the 3D Ising Model \\\vspace{.1in}}
\end{center}
}
\vspace{.2in}
\begin{center}
{\bf 
Filip Kos$^{a}$, 
David Poland$^{a}$,
David Simmons-Duffin$^{b}$} 
\\
\vspace{.2in} 
$^a$ {\it  Department of Physics, Yale University, New Haven, CT 06520}\\
$^b$ {\it School of Natural Sciences, Institute for Advanced Study, Princeton, New Jersey 08540}
\end{center}

\vspace{.2in}

\begin{abstract}
We study the conformal bootstrap for systems of correlators involving non-identical operators. The constraints of crossing symmetry and unitarity for such mixed correlators can be phrased in the language of semidefinite programming. We apply this formalism to the simplest system of mixed correlators in 3D CFTs with a $\mathbb{Z}_2$ global symmetry. For the leading $\mathbb{Z}_2$-odd operator $\s$ and $\mathbb{Z}_2$-even operator $\e$, we obtain numerical constraints on the allowed dimensions $(\De_\s, \De_\e)$ assuming that $\s$ and $\e$ are the only relevant scalars in the theory. These constraints yield a small closed region in $(\De_\s, \De_\e)$ space compatible with the known values in the 3D Ising CFT.
\end{abstract}

\newpage

\tableofcontents

\newpage

\section{Introduction}
\label{sec:intro}

The conformal bootstrap~\cite{Polyakov:1974gs} in $D>2$ dimensions has produced remarkable results, including numerical bounds on operator dimensions and OPE coefficients~\cite{Rattazzi:2008pe,Rychkov:2009ij,Caracciolo:2009bx,Poland:2010wg,Rattazzi:2010gj,Rattazzi:2010yc,Vichi:2011ux,Poland:2011ey,Rychkov:2011et,ElShowk:2012ht,Liendo:2012hy,ElShowk:2012hu,Beem:2013qxa,Kos:2013tga,Gliozzi:2013ysa,El-Showk:2013nia,Alday:2013opa,Gaiotto:2013nva,Bashkirov:2013vya,Berkooz:2014yda,El-Showk:2014dwa,Gliozzi:2014jsa,Nakayama:2014lva,Alday:2014qfa}, analytical constraints~\cite{Heemskerk:2009pn,Heemskerk:2010ty,Fitzpatrick:2012yx,Komargodski:2012ek,Beem:2013sza,Fitzpatrick:2014vua,Beem:2014kka}, and recently a precise conjecture for the $\mathbb{Z}_2$-even spectrum of the 3D Ising model~\cite{El-Showk:2014dwa}. However, all previous studies have focused on a single four-point function $\<\f\f\f\f\>$ containing identical operators (sometimes in a nontrivial global symmetry representation). It is extremely important to ask how other correlators, such as $\<\f \f \f^2 \f^2\>$ or more generally $\<\f_1 \f_2 \f_3 \f_4\>$ for different operators $\f_i$, additionally constrain the space of CFTs.

An immediate complication is that the unitarity properties of such mixed correlators are more intricate because coefficients in the conformal block expansion are not necessarily positive. In section~\ref{sec:bootstrap}, we describe how these unitarity properties can be captured by a semidefinite program with a continuously infinite number of constraints (as opposed to the linear programs that arise in the single correlator case).  Applying methods introduced in~\cite{Poland:2011ey,Kos:2013tga}, we rewrite this as a higher-dimensional semidefinite program with a finite number of constraints, which can be solved on a computer.\footnote{There are still a discretely infinite number of constraints, labeled by spins $\ell$ that can appear in the OPE.  However in practice, it is sufficient to include a large but finite number of spins, see appendix~\ref{app:sdpa}.}  In section~\ref{sec:bootstrapforscalars}, we specialize our discussion to the case of scalar correlators with a $\mathbb{Z}_2$ global symmetry.  

To formulate our semidefinite program, we need approximations for conformal blocks as rational functions of the exchanged operator dimension $\De$. Such approximations follow from a rapidly convergent expansion for the blocks as a sum over poles $1/(\De-\De_i)$. We describe this expansion in section~\ref{sec:rational}, generalizing the results of~\cite{Kos:2013tga} to non-equal external operator dimensions. Our expressions give higher dimensional analogs of a recursion relation for Virasoro conformal blocks developed by Alyosha Zamolodchikov in~\cite{Zamolodchikov:1985ie,Zamolodchikov:1987}.

In section~\ref{sec:results} we apply our formalism to numerically study operator dimensions in 3D CFTs with a $\mathbb{Z}_2$ global symmetry, a class of theories that includes the 3D Ising model.  We focus on the system of four-point functions $\{\<\s\s\s\s\>, \<\s\s\e\e\>, \<\e\e\e\e\>\}$ containing the lowest $\Z_2$-odd scalar $\s$ and $\Z_2$-even scalar $\e$, and ask: {\it What are the allowed dimensions $(\De_\s,\De_\e)$ assuming that $\s$ and $\e$ are the only relevant scalars in the CFT?}

The existence of only two relevant scalars is an obvious experimental fact about the 3D Ising CFT --- it follows from the observation that the phase diagram of water is two-dimensional.  Nevertheless, despite the mild assumptions of $\Z_2$ symmetry and two relevant scalars, we find a striking result: the dimensions $(\De_\s,\De_\e)$ are almost uniquely fixed!  The allowed region is a tiny sliver around $\De_\s=0.51820(14)$ and $\De_\e=1.4127(11)$, in agreement with the most precise Monte-Carlo simulations~\cite{Hasenbusch:2011yya}, and also the $c$-minimization conjecture of \cite{El-Showk:2014dwa}.  Indeed, our results give strong support for $c$-minimization --- support that can be strengthened with further numerical work.

It is plausible that with arbitrary computational power, the constraints we study are strong enough to uniquely determine the spectrum of the 3D Ising CFT.  In particular, our results support the conjecture that the 3D Ising CFT is the only $\Z_2$-symmetric 3D CFT with exactly two relevant operators, giving a higher-dimensional example of {\it critical universality}~\cite{Kadanoff:1971pc,Stanley:1999zz}.  

Because our study includes the correlator $\<\s\s\e\e\>$, we can additionally learn about $\Z_2$-odd operators (appearing in the $\s\x\e$ OPE), which were not accessible in previous bootstrap studies.  In section~\ref{sec:results} we also compute an upper bound $\De_{\s'}\leq  5.41(1)$, where $\s'$ is the second-lowest dimension $\Z_2$-odd scalar.  A precise determination of the complete $\Z_2$-odd spectrum of the 3D Ising model is a fascinating problem that we leave for future work.  We describe other important future directions in section~\ref{sec:discussion}.

\section{Bootstrapping Mixed Correlators and Semidefinite Programming}
\label{sec:bootstrap}

\subsection{What is New About Mixed Correlators?}
\label{sec:whatsnew}

It was shown in \cite{Rattazzi:2008pe} that the bootstrap constraints for a four-point function of identical scalars $\<\f\f\f\f\>$ can be transformed into a system of linear inequalities.  Studying solutions to these inequalities leads to bounds on CFT data.  By contrast, the bootstrap constraints for mixed correlators $\<\f_1\f_2\f_3\f_4\>$ cannot be written in terms of linear inequalities --- rather the mixed correlator problem is intrinsically quadratic.\footnote{In fact, as we'll see shortly, one has a quadratic problem even for identical operators with spin or identical operators in large global symmetry representations.}  In this section, we describe how this quadratic problem arises in a simple case, and how the procedure of \cite{Rattazzi:2008pe} can be modified to solve it.

\subsubsection{Review of the Bootstrap Argument for Identical Scalars}

Let us first recall the original bootstrap argument of \cite{Rattazzi:2008pe} for a four-point function of identical real scalars $\<\f(x_1)\f(x_2)\f(x_3)\f(x_4)\>$.  Using the OPE, we can write the four-point function as a sum over conformal blocks
\be
\<\contraction{}{\f}{(x_1)}{\f}\f(x_1)\f(x_2)\contraction{}{\f}{(x_3)}{\f}\f(x_3)\f(x_4)\> &=& \frac{1}{x_{12}^{2\De_\f}x_{34}^{2\De_\f}}\sum_{\cO\in\f\x\f} \l_{\f\f\cO}^2 g_{\De,\ell}(u,v).
\ee
Here, $u=\frac{x_{12}^2 x_{34}^2}{x_{13}^2 x_{24}^2}, v=\frac{x_{23}^2x_{14}^2}{x_{13}^2 x_{24}^2}$ are conformal cross-ratios, $\cO$ runs over real primary operators appearing in the $\f\x\f$ OPE, $\De=\dim\,\cO$, and $\ell=\mathrm{spin}\,\cO$.  The OPE coefficients $\l_{\f\f\cO}$ are real by unitarity, implying that their squares $\l_{\f\f\cO}^2$ are positive.

The four-point function should be independent of how we pair the operators to perform the OPE.  Specifically, swapping $1\leftrightarrow 3$, we find the crossing equation
\be
v^{\De_\f}\sum_{\cO}\l_{\f\f\cO}^2 g_{\De,\ell}(u,v) &=& u^{\De_\f}\sum_\cO \l_{\f\f\cO}^2 g_{\De,\ell}(v,u).
\ee
Grouping terms that multiply $\l_{\f\f\cO}^2$, we obtain a sum rule with positive coefficients
\be
\label{eq:sumrule}
\sum_{\cO} \l_{\f\f\cO}^2 F_{\De,\ell}(u,v) &=& 0,\\
F_{\De,\ell}(u,v)&\equiv& v^{\De_\f}g_{\De,\ell}(u,v) - u^{\De_\f}g_{\De,\ell}(v,u)\label{eq:definitionofF}.
\ee

Positivity of the coefficients in (\ref{eq:sumrule}) is the key property that leads to universal bounds on CFT data, {\it without having to know the precise details of the operators $\cO$ entering the OPE}.  Let us first assume the dimensions $\De$ and spins $\ell$ lie in some specified range.  For example, we might assume that all scalars have dimension larger than some $\De_0$.  Consider linear functionals $\a$ acting on functions of $u,v$, and suppose there exists an $\a$ satisfying the conditions
\be
\label{eq:linearprogram}
\a(F_{\De,\ell})&\geq& 0\textrm{ for all $\De,\ell$ in the spectrum, and}\nn\\
\a(F_{0,0})&=&1.
\ee
If such an $\a$ exists, the sum rule (\ref{eq:sumrule}) cannot be satisfied with any choice of operators $\cO$, and the hypothetical CFT is ruled out.

We can always express $\a$ in terms of a basis of functionals, for example derivatives around the crossing-symmetric point,
\be
\label{eq:ansatzforalpha}
\a: F \mto \sum_{m,n}\left. a_{mn}\ptl_z^m\ptl_{\bar z}^nF(z,\bar z)\right|_{z=\bar z = \frac 1 2},
\ee
where $z,\bar z$ are defined by $u=z\bar z$, $v=(1-z)(1-\bar z)$.  Then, (\ref{eq:linearprogram}) becomes a set of linear inequalities (and one affine equality) for the coefficients $a_{mn}$.

\subsubsection{Applying the Bootstrap Argument to a Mixed Correlator}

Let us try to apply this procedure to $\<\f_1(x_1)\f_2(x_2)\f_2(x_3)\f_1(x_4)\>$, where $\f_1$ and $\f_2$ are different scalar primaries.  Equating conformal block expansions in the $(12)\to(34)$ and $(23)\to(14)$ channels, we have
\be
v^{\De_2} \sum_{\cO} |\l_{12\cO}|^2 g_{\De,\ell}^{\De_{12},\De_{21}}(u,v) = u^{\frac{\De_1+\De_2}{2}} \sum_{\cO} \l_{22\cO} \l_{11\cO} g^{0,0}_{\De,\ell}(v,u),
\label{eq:nonpositivecrossing}
\ee
where $g_{\De,\ell}^{\De_{ij},\De_{kl}}(u,v)$ is a conformal block for scalars with possibly unequal dimensions, $\De_{ij}\equiv \De_i-\De_j$, and $\l_{ij\cO}$ denotes the OPE coefficient of $\cO\in \f_i\x\f_j$.  We will often abbreviate $g^{0,0}_{\De,\ell}(u,v)$ as $g_{\De,\ell}(u,v)$.

The left-hand side of (\ref{eq:nonpositivecrossing}) has manifestly positive coefficients $|\l_{12\cO}|^2$.  However, on the right-hand side there is no a-priori relation between $\l_{22\cO}$ and $\l_{11\cO}$, so their product can have either sign.  Consequently, we cannot simply apply linear functionals to both sides and derive conclusions about the allowed spectrum.  We must modify the bootstrap logic above.

To obtain some kind of positivity condition, we can combine the crossing equation for $\<\f_1\f_2\f_2\f_1\>$ with crossing equations for $\<\f_1\f_1\f_1\f_1\>$ and $\<\f_2\f_2\f_2\f_2\>$ into one equation
\be
\sum_{\cO}\begin{pmatrix}
\l_{11\cO} & \l_{22\cO}
\end{pmatrix}
\begin{pmatrix}
F^{(11)}_{\De,\ell} & F^{(12)}_{\De,\ell}\\
F^{(21)}_{\De,\ell} & F^{(22)}_{\De,\ell}\\
\end{pmatrix}
\begin{pmatrix}
\l_{11\cO}\\
\l_{22\cO}
\end{pmatrix}
+\textrm{$\l_{12\cO}^2$ terms} &=& 0,
\label{eq:examplematrixcrossingeq}
\ee
where the $F_{\De,\ell}^{(ij)}$ are combinations of conformal blocks analogous to (\ref{eq:definitionofF}), and we have suppressed their $u,v$ dependence for brevity.  The quantity multiplying $\l_{11\cO}$ and $\l_{22\cO}$ above is a $2\x2$ matrix of functions of $u, v$.  Formally, it is an element of $\R^{2\x2}\otimes \cF$, where $\cF$ is the space of functions of $u,v$ in the region where both conformal block expansions converge.\footnote{The convergence region $\cF$ includes a finite open neighborhood of the point $z=\bar z = \frac 1 2$ \cite{Pappadopulo:2012jk}.}  The right-hand side of (\ref{eq:nonpositivecrossing}) contributes to off-diagonal elements of this matrix.

We can now consider linear functionals $\a$ acting on functions of $u,v$
\be
\a:\cF \to \R,
\ee
such that
\be
\begin{pmatrix}
\a(F^{(11)}_{\De,\ell}) & \a(F^{(12)}_{\De,\ell})\\
\a(F^{(21)}_{\De,\ell}) & \a(F^{(22)}_{\De,\ell})\\
\end{pmatrix}\in\R^{2\x 2} \textrm{ is a positive semidefinite matrix},
\label{eq:examplepositivesemidefiniteness}
\ee
and $\a$ is additionally positive acting on each $\l_{12\cO}^2$ term in (\ref{eq:examplematrixcrossingeq}).  Applying $\a$ to both sides, we again have termwise positivity and the bootstrap logic can proceed.

An optimization problem that includes positive semidefiniteness constraints of the form (\ref{eq:examplepositivesemidefiniteness}) is a {\it semidefinite program}, as opposed to the {\it linear program}~(\ref{eq:linearprogram}) that appears in the case of identical operators.\footnote{Because of the infinite number of constraints (one for each $\De,\ell$), one technically has a {\it semi-infinite} program in the identical operators case.  We will not bother with this distinction.  We do not know the correct terminology for a semidefinite program with an infinite number of constraints, as in the case of mixed operators.}  Semidefinite programming has appeared in the conformal bootstrap before: it was applied to 4D CFTs in \cite{Poland:2011ey}, and later extended to arbitrary spacetime dimensions in \cite{Kos:2013tga}.  However, its appearance here is qualitatively different.  In \cite{Poland:2011ey,Kos:2013tga}, semidefinite programming was a useful trick for efficiently encoding the infinite number of constraints $\a(F_{\De,\ell})\geq 0$ (one for each $\Delta$ and $\ell$).  This trick is not strictly necessary, and alternative methods have also been successful, for example the discretization of $\De$ in \cite{Rattazzi:2008pe} and the modified simplex algorithm in \cite{El-Showk:2014dwa}.  By contrast, the appearance of semidefinite programming in (\ref{eq:examplepositivesemidefiniteness}) is unavoidable, stemming from the intrinsically quadratic nature of the crossing constraints.\footnote{It is possible to approximate the semidefiniteness constraint with a finite number of linear constraints by approximating the cone of semidefinite matrices as a polytope, as explored in \cite{SlavaUnpublished}.}

In this work, we will combine the semidefinite programming trick of \cite{Poland:2011ey,Kos:2013tga} with the novel appearance of semidefinite programming in~(\ref{eq:examplepositivesemidefiniteness}).  Thus, semidefinite programming appears in two ways: one optional, the other obligatory.  An alternative approach that may be fruitful would be to adapt the modified simplex algorithm of \cite{El-Showk:2014dwa} to work with semidefiniteness constraints.\footnote{Semidefinite programs enjoy a duality similar to the linear programming duality underlying the primal simplex method in \cite{El-Showk:2014dwa}.}

\subsection{Spin and Global Symmetry Representations}

Before describing our construction in detail, let us note that the appearance of semidefinite programming is generic in the conformal bootstrap, and previously considered problems are special cases where it can be avoided.  Semidefinite programming appears whenever a conformal block expansion has coefficients with indefinite sign.

An important example is a four-point function of operators with spin, where multiple structures can appear in the OPE.  For example, let $J^\mu(x)$ be a conserved current in a 3D CFT.  A primary operator $\cO^{\mu_1\cdots\mu_\ell}$ with even spin $\ell$ can appear in the $J\x J$ OPE with two different parity-even tensor structures \cite{Giombi:2011rz,Costa:2011mg,Costa:2011dw},
\be
J^\mu(x)J^\nu(0) &\sim& \l^1_{JJ\cO} t_1^{\mu\nu}{}_{\mu_1\dots\mu_\ell}(x)\cO^{\mu_1\dots\mu_\ell}(0)+\l^2_{JJ\cO} t_2^{\mu\nu}{}_{\mu_1\dots\mu_\ell}(x)\cO^{\mu_1\dots\mu_\ell}(0)+\mathrm{descendants}.\nn\\
\ee
The tensors $t_1(x)$ and $t_2(x)$ are fixed by conservation, symmetry under exchanging the $J$'s, and conformal invariance.  Each tensor structure has an independent OPE coefficient $\l^1_{JJ\cO},\l^2_{JJ\cO}$, and thus the conformal block expansion of $\<J^\mu J^\nu J^\rho J^\s\>$ contains terms proportional to $\l^1_{JJ\cO}\l^2_{JJ\cO}$ (which can have either sign).

Another example is a four-point function of operators in representations $\br_i$ of a global symmetry group $G$, where the tensor product $\br_i\otimes\br_j$ contains irreducible representations with nontrivial multiplicity.  The fact that bootstrap conditions become semidefiniteness constraints in this case was first noticed in \cite{Rattazzi:2010yc}.

For instance, consider $\<\f_{\br}(x_1)\f_{\br}(x_2)\f_{\br}^\dag(x_3)\f_{\br}^\dag(x_4)\>$, where $\f_{\br}$ is a scalar operator transforming in the representation $\br$, and $\f^\dag_{\br}$ transforms in the dual representation $\bar \br$.  Suppose $\cO_{\bs,\ell}\in\f_\br\x\f_\br$ is a spin-$\ell$ operator transforming in the representation $\bs$ of $G$.  The three-point function
$
\<\f_\br\f_\br\cO_{\bs,\ell}^\dag\>
$
must be proportional to an invariant tensor $t_\cO$ of $G$.  Specifically,
\be
\<\f_\br\f_\br\cO_{\bs,\ell}^\dag\> &\propto& t_\cO \in \left\{
\begin{array}{ll}
(\Sym^2\br \otimes \bar\bs)^G & \textrm{if $\ell$ is even,}\\
(\we^2\br \otimes \bar\bs)^G & \textrm{if $\ell$ is odd.}
\end{array}
\right.
\ee
where $\Sym^2$ and $\we^2$ denote symmetric and antisymmetric tensor squares, and $(\cdot)^G$ denotes the $G$-invariant subspace.  The space of such invariant tensors may be multidimensional: its dimension counts the multiplicity of $\bs$ in the decomposition of $\Sym^2 \br$ and $\we^2 \br$ into irreducibles. If so, we can expand $t_\cO$ in a basis of invariant tensors $t_i$, each with an independent coefficient
\be
t_\cO=\l^1_{\f\f\cO} t_1+\l^2_{\f\f\cO} t_2+\dots.
\ee
The conformal block expansion of our four-point function in the $(12)\to(34)$ channel can then contain products $\l_{\f\f\cO}^{i}\l_{\f\f\cO}^{j*}$, which are not necessarily positive.

Previous bootstrap studies of CFTs with global symmetries have focused on small representations: either $G=\SO(n)$ with $\br$ the vector representation, or $G=\SU(n)$ with $\br$ the fundamental representation.  In each of these cases, the spaces $\Sym^2\br$, $\we^2\br$, and $\br\otimes \bar\br$ (in the other channel) decompose into irreducibles with multiplicity at most 1.

However, it is easy to find examples with higher multiplicities.  For instance, take $G=\SU(n)$, $n \ge 3$, and let $\br$ to be the largest irreducible representation in $\Sym^2\mathrm{Ad}_G$, of dimension $\frac 1 4 n^2(n-1)(n+3)$.  In this case,
\be
\Sym^2 \br = 2\,\br\oplus\dots
\ee
so that there are two independent OPE coefficients for an operator $\cO_{\br,\ell}$ with even spin.  The three-point structures for this example are written explicitly in appendix~\ref{app:threeptglobalexample}.

\subsection{General Semidefinite Programs for the Bootstrap}
\label{sec:generalsdpbootstrap}

In the previous subsection, we saw examples of different ways that semidefinite programming can arise in the bootstrap. Now let us generalize these examples and show how the generic statement of crossing symmetry and unitarity can be phrased as a semidefinite program.  The discussion here is somewhat abstract.  In section~\ref{sec:bootstrapforscalars}, we will specialize to the case of interest for the remainder of this work.

Consider a CFT whose symmetry group $H$ is a product of the conformal group and global symmetry groups.\footnote{In our discussion, we focus on bosonic operators in non-SUSY CFTs.  The generalization of this section to fermionic operators and/or superconformal theories is straightforward.}  Primary operators $\cO_i$ transform in unitary irreducible  representations $R_i$ of $H$.  Let $H_0$ be the isotropy subgroup of $H$ (the group of symmetries that fix a spacetime point), which is generated by Lorentz transformations, dilatations, special conformal transformations, and global symmetry transformations.  The representations $R_i$ are induced from finite dimensional representations $R_{i0}$ of $H_0$.  Hence, each $\cO_i$ carries an index $a_i$ for $R_{i0}$.  For example, an uncharged spin-1 operator $J^\mu$ has $a_i=\mu$, a Lorentz index.

The four-point functions of the theory are given by
\be
\cG_{ijkl}^{a_ia_ja_ka_l}(x_1,x_2,x_3,x_4) &\equiv& \<\cO_i^{a_i}(x_1)\cO_j^{a_j}(x_2)\cO_k^{a_k}(x_3)\cO_l^{a_l}(x_4)\>.
\ee
For each set of representations $R_i,R_j,R_k,R_l$, the four-point function can be expanded in a finite basis of four-point structures $I^{a_ia_ja_ka_l}_f(x_i)$, times functions $G_{ijkl}^{f}(u,v)$ of conformal cross-ratios,
\be
\label{eq:fourpointfunctiondecompostition}
\cG_{ijkl}^{a_ia_ja_ka_l}(x_1,x_2,x_3,x_4) &=& \sum_f G_{ijkl}^{f}(u,v) I^{a_ia_ja_ka_l}_f(x_i).
\ee
The number of four-point structures depends on the representations $R_i,R_j,R_k,R_l$.  For example, when all the operators are scalars there is a single four-point structure. Its definition is ambiguous up to multiplication by conformal cross ratios $u,v$. We choose
\be
I(x_i) &=& \frac{1}{x_{12}^{\De_i + \De_j} x_{34}^{\De_k+\De_l}} \left(\frac{x_{24}}{x_{14}}\right)^{\De_{ij}} \left(\frac{x_{14}}{x_{13}}\right)^{\De_{kl}},
\label{eq:fourptforscalars}
\ee
where $x_{ij}=|x_i-x_j|$ and $\De_{ij}\equiv \De_i-\De_j$.  

Crossing symmetry is the statement that (for bosonic operators) swapping $(i,a_i,x_1)\leftrightarrow(k,a_k,x_3)$ leaves the four-point function unchanged,\footnote{There are, of course, other crossing equations for other permutations of the four operators.  The permutations $1\leftrightarrow 2$ and $3\leftrightarrow 4$ typically give constraints that are simple to analyze.  The permutation $1\leftrightarrow 3$ above gives a new nontrivial constraint.  All other permutations can be obtained as compositions of these.}
\be
\cG_{ijkl}^{a_ia_ja_ka_l}(x_1,x_2,x_3,x_4)&=&\cG_{kjil}^{a_ka_ja_ia_l}(x_3,x_2,x_1,x_4)\nn\\
&=& \sum_{f'}G^{f'}_{kjil}(v,u)(I_{f'}^{a_ka_ja_ia_l}(x_i)|_{x_1\leftrightarrow x_3})\nn\\
&=& \sum_{f,f'}G^{f'}_{kjil}(v,u)S_{f'}{}^f(u,v) I_{f}^{a_ia_ja_ka_l}(x_i),
\ee
where $S_{f'}{}^f(u,v)$ is a finite-dimensional matrix defined by
\be
\sum_f S_{f'}{}^f(u,v) I_{f}^{a_ia_ja_ka_l}(x_i) &=& I_{f'}^{a_ka_ja_ia_l}(x_i)|_{x_1\leftrightarrow x_3}.
\ee
For instance, for scalar operators, using (\ref{eq:fourptforscalars}) we have
\be
S  &=& \frac{x_{12}^{\De_i + \De_j} x_{34}^{\De_k+\De_l}}{x_{23}^{\De_j + \De_k} x_{14}^{\De_i+\De_l}} \left(\frac{x_{24}}{x_{34}}\right)^{\De_{kj}} \left(\frac{x_{34}}{x_{13}}\right)^{\De_{il}} \left(\frac{x_{24}}{x_{14}}\right)^{\De_{ji}} \left(\frac{x_{14}}{x_{13}}\right)^{\De_{lk}}=\frac{u^{\frac{\De_i+\De_j}{2}}}{v^{\frac{\De_j+\De_k}{2}}}.
\ee

We should think of $G_{ijkl}^f(u,v)$ as a vector with indices given by $ijkl,f,u,$ and $v$.  The matrix $S$ acts on the $f$ and $u,v$-indices.  For convenience, let us additionally define the operator $T:G_{ijkl}^f(u,v)\mto G_{kjil}^f(u,v)$ which swaps $i\leftrightarrow k$ and also $U:G(u,v)\mto G(v,u)$ which swaps $u\leftrightarrow v$.  Crossing symmetry then states
\be
0 &=& (1-STU)(G).
\ee

As usual in the bootstrap, to constrain solutions to crossing symmetry, we can look at linear functionals acting on the crossing equation.  A linear functional acting on $G$ has the form
\be
\a(G) &=& \sum_{ijkl;f}\a^{ijkl}_{f}(G^f_{ijkl}) ,
\ee
where each $\a^{ijkl}_{f}$ acts on functions of two variables, as in (\ref{eq:ansatzforalpha}).  The dual of the crossing equation is the statement that
\be
0 &=& \a((1-STU)(G))\nn\\
&=& \sum_{ijkl;f} (\a^{ijkl}_{f} - \a^{kjil}_{f'}\circ (S_{f}{}^{f'}U))(G_{ijkl}^f)
\label{eq:dualizedcrossingequation}
\ee
for all $\a$.

The final ingredient is the conformal block expansion for the functions $G^f_{ijkl}$.  For an operator $\cO_p$ appearing in the OPE $\cO_i\x\cO_j$, there are in general several three-point structures $t_{m}$ that can appear, and each structure has an associated OPE coefficient $\l^{(m)}$,
\be
\cO_i^{a_i}(x) \cO_j^{a_j}(0) &=& \sum_p\sum_m \l_{ijp}^{(m)} t^{a_i a_j}_{m}{}_{a_p}(x,\ptl)\cO_p^{a_p}(0).
\ee
Consequently, conformal blocks in the $ij\to kl$ channel are labeled by pairs of three-point structures $(m,n)$,
\be
\label{eq:conformalblockexpansion}
G_{ijkl}^f &=& \sum_{p,m,n} \l_{ijp}^{(m)}\l_{klp}^{(n)} g^{R_iR_jR_kR_l;f}_{R_p(m,n)}(u,v).
\ee
The conformal blocks $g_{R_p(m,n)}^{R_iR_jR_kR_l;f}$ depend on the representations (dimensions, spins, and global symmetry charges) of the external and exchanged operator, together with the three-point structures $(m,n)$ and four-point structure $f$.

Plugging the conformal block expansion (\ref{eq:conformalblockexpansion}) into the dualized crossing equation (\ref{eq:dualizedcrossingequation}) gives
\be
\label{eq:dualizedcrossingequationwithconformalblocks}
0 &=& \sum_p\sum_{ijm}\sum_{kln} \l_{ijp}^{(m)}\l_{klp}^{(n)} \p{\a_f^{ijkl}(g_{R_p(m,n)}^{R_iR_jR_kR_l;f})-\a_{f'}^{kjil}(S_{f}{}^{f'}Ug_{R_p(m,n)}^{R_iR_jR_kR_l;f})}.
\ee
All repeated indices are summed, but we have chosen to indicate some of the sums explicitly.  Let us think of $\l_{ijp}^{(m)}$ as a vector $\vec{\l}_p$ with indices $ij,m$.  For each $p$, the quantity in parentheses is then a matrix $A_p$ with left indices $ij,m$ and right indices $kl,n$, and the above equation can be written
\be
0 &=& \sum_p \vec \l_p^T A_p \vec \l_p.
\label{eq:finallysemidefinite}
\ee

From (\ref{eq:finallysemidefinite}), it is clear that the crossing equation can be studied using semidefinite programming.  Specifically, we can make an assumption about the CFT spectrum, and then search for $\a$ such that
\be
A_0 &\succ& 0,\quad\textrm{where $\cO_0=\mathbf{1}$ is the unit operator,}\nn\\
A_p &\succeq& 0,\quad \textrm{for all $p$ which can appear in the spectrum.}
\ee
If such an $\a$ exists, the assumed spectrum is ruled out.  As usual in the bootstrap, the advantage of studying this dual formulation of crossing symmetry is that one can make progress by restricting to a finite-dimensional space of $\a$'s.  By contrast, truncating the operator spectrum itself can introduce approximations that are difficult to control.

\subsection{SDP${}^2$}
\label{sec:sdp2}

The semidefinite programs described in previous sections have an infinite number of positivity constraints~(\ref{eq:examplepositivesemidefiniteness}) --- one for each $\De$ and $\ell$ (and more generally for each conformal representation) that can appear in the spectrum.  To encode these constraints on a computer, we write them in terms of a finite amount of data using a trick from \cite{Poland:2011ey,Kos:2013tga}.

Let us briefly review this trick in the case of identical operators.  The positivity constraints (\ref{eq:linearprogram}) are linear inequalities of the form
\be
\label{eq:identicalopspositivity}
\a(F_{\De,\ell})\geq 0\quad \textrm{for}\quad\De\geq \De_{\ell}^\mathrm{min},
\ee
where $\De_\ell^\mathrm{min}$ is an $\ell$-dependent lower bound on the dimension (e.g., the unitarity bound).
We take $\a$ to be of the form~(\ref{eq:ansatzforalpha}): a sum of derivatives with respect to $z,\bar z$ around the crossing symmetric point $z=\bar z = 1/2$.

The trick proceeds by first rewriting our inequalities as positivity of polynomials.  This is possible because there exists a systematic approximation
\be
\label{eq:approxforderivatives}
\ptl_z^m\ptl_{\bar z}^n F_{\De,\ell}(1/2,1/2) &\approx& \chi_\ell(\De) P^{(m,n)}_\ell(\De),
\ee
where $\chi_\ell(\De)$ is a positive function and $P^{(m,n)}_\ell(\De)$ are polynomials in $\De$.\footnote{Such approximations were first discovered for 2D and 4D theories in \cite{Poland:2011ey}.  Their existence in arbitrary dimensions was shown to follow from representation theory of the conformal algebra in \cite{Kos:2013tga}.  We review this argument in section~\ref{sec:rational}.} Dividing~(\ref{eq:identicalopspositivity}) by $\chi_\ell(\De)$, we obtain
\be
\label{eq:polynomialconstraints}
\sum_{m,n} a_{mn} P^{(m,n)}_\ell(\De_\ell^\mathrm{min}+x) \geq 0\quad\textrm{for}\quad x\geq 0,
\ee
where we defined $x=\De-\De_\ell^\mathrm{min}$.
The left hand side is now a polynomial in $x$.  The degree of this polynomial grows logarithmically with the accuracy of the approximation~(\ref{eq:approxforderivatives}).  

A classic theorem \cite{PolyaSzego:1976} states that a polynomial $P(x)$ is positive on the positive real axis,
\be
P(x)\geq 0\quad\textrm{for}\quad x\in[0,\oo)
\ee
if and only if
\be
\label{eq:sumofsquares}
P(x) &=& a(x)+x b(x),
\ee
where $a(x)$ and $b(x)$ are sums of squares of polynomials.  That is, $a(x)=\sum_t p_t(x)^2$ and similarly for $b(x)$.  

Let $[x]_d=(1,x,\dots,x^d)^T$ be a vector of monomials up to degree $d$ and define the $(d+1)\x(d+1)$ matrix $Q_d(x)\equiv [x]_d [x]_d^T$.  It is easy to show that any degree-$2d$ sum of squares is of the form $\Tr (A Q_d(x))$, where $A\succeq 0$ is a positive semidefinite matrix.  Consequently, (\ref{eq:sumofsquares}) is equivalent to
\be
\label{eq:polynomialpositivityintermsofmatrices}
P(x)=\Tr(A Q_{d_1}(x)) + x\Tr(B Q_{d_2}(x))\quad\textrm{with}\quad A,B\succeq 0.
\ee
Here $d_1=\lfloor \frac 1 2\deg P\rfloor$, $d_2=\lfloor\frac 1 2 (\deg P - 1)\rfloor$, and ``$\succeq$" means ``positive semidefinite." Using~(\ref{eq:polynomialpositivityintermsofmatrices}), we can write the constraints (\ref{eq:polynomialconstraints}) in terms of positive semidefinite matrices $A_\ell,B_\ell$ with linear relations between the elements of $A_\ell,B_\ell$ and the coefficients $a_{mn}$.  This efficiently encodes an infinite number of constraints in terms of finite matrices.

Now let us consider the case of distinct operators, with constraints of the form (\ref{eq:examplepositivesemidefiniteness}),
\be
\begin{pmatrix}
\a(F^{(11)}_{\De,\ell}) & \dots & \a(F^{(1N)}_{\De,\ell})\\
\vdots & \ddots & \vdots\\
\a(F^{(N1)}_{\De,\ell}) & \dots & \a(F^{(NN)}_{\De,\ell})\\
\end{pmatrix}\succeq 0 \quad\textrm{for} \quad\De\geq \De_\ell^{\mathrm{min}}.
\label{eq:distinctpositivity}
\ee
Here, $F^{(ij)}_{\De,\ell}$ are combinations of conformal blocks, and  $\a$ is again a sum of derivatives with respect to $z,\bar z$.  Similarly to~(\ref{eq:approxforderivatives}), a systematic positive-times-polynomial approximation for each entry of the above matrix exists,
\be
\ptl_z^m\ptl_{\bar z}^n F^{(ij)}_{\De,\ell}(1/2,1/2) &\approx& \chi_\ell(\De) P^{(ij;m,n)}_\ell(\De).
\ee 
Crucially, the positive function $\chi_\ell(\De)$ is independent of the matrix indices $i,j$.  Dividing by $\chi_\ell(\De)$, (\ref{eq:distinctpositivity}) is then equivalent to
\be
\sum_{m,n}a_{mn}
\begin{pmatrix}
P^{(11;m,n)}_\ell(x) &\dots& P_\ell^{(1N;m,n)}(x)\\
\vdots & \ddots &\vdots\\
P^{(N1;m,n)}_\ell(x) & \dots & P_\ell^{(NN;m,n)}(x)
\end{pmatrix}\succeq 0 \quad\textrm{for}\quad x\geq 0,
\label{eq:matrixpolynomialconstraints}
\ee
where $P_\ell^{(ij;m,n)}(x)$ are polynomials and $x=\De-\De^\mathrm{min}_\ell$.

Analogously to the case of positive polynomials, an $N\x N$ matrix polynomial that is positive on the positive real axis, such as the one in (\ref{eq:matrixpolynomialconstraints}), can be written as \cite{HanselkaUnpublished}
\be
a(x) + x b(x),
\ee
where $a(x)$ and $b(x)$ are sums of squares of matrix polynomials, i.e. $a(x) = \sum_t p_t^T(x) p_t(x)$, and similarly for $b(x)$.\footnote{We thank Christoph Hanselka and Markus Schweighofer for providing a proof of this claim.} Let the highest degree of the polynomials $p_t$ be $d_1$, and write the matrix polynomials $p_t(x)$ as
\be
p_t(x) = \sum_{\rho=0}^{d_1} p_{t;\rho} x^\rho,
\ee
where $p_{t;\rho}$ are $M\times N$ matrices.  Any matrix polynomial $a(x)$ can be written as
\be
a(x) = \Tr_{\R^{d_1+1}}(A(Q_{d_1}(x)\otimes \mathbf{1})),
\ee
where $A$ is an $N(d_1+1)\x N(d_1+1)$ matrix acting on $\R^{d_1+1}\otimes \R^{N}$.  $Q_{d_1}(x)$ acts only on the first tensor factor $\R^{d_1+1}$ and the trace is over this factor, leaving an $N\x N$ matrix that depends on $x$.
In terms of components, the elements of $a$ are given by
\be
a^{ij}(x) = [x]_{d_1}^T A^{ij} [x]_{d_1} = \sum_{\rho\sigma} A_{\rho \sigma}^{ij} x^{\rho} x^{\sigma}.
\ee

If $a(x)$ is a sum of matrix squares, then $A$ is positive semidefinite on the space $\R^{d_1+1}\otimes \R^N$. To see this, note that for an arbitrary $N(d_1+1)$-vector $\lambda_{\rho}^{i}$ we have
\be
(\lambda, A\lambda) = \sum_{ij,\rho\sigma} A_{\rho\sigma}^{ij} \lambda_{\rho}^{i} \lambda_{\sigma}^j =\sum_{t,ijk,\rho\sigma} p_{t;\rho}^{ki} p_{t;\sigma}^{kj} \lambda_{\rho}^{i} \lambda_{\sigma}^j = \sum_{t,ik,\rho} (p_{t;\rho}^{ki}\lambda_{\rho}^{i} )^2 \ge 0.
\ee
The converse also holds: if $A$ is positive semidefinite, then $a(x)$ is a sum of matrix squares, as can be seen by writing $A$ as a sum of outer-products.
Therefore, the matrix polynomials of (\ref{eq:matrixpolynomialconstraints}) can be written in a way analogous to (\ref{eq:polynomialpositivityintermsofmatrices}):
\be
\sum_{m,n}a_{mn} P^{(ij;m,n)}_\ell(x)  &=& [x]_{d_1}^T A^{ij} [x]_{d_1} + x [x]_{d_2}^T B^{ij} [x]_{d_2},\nn\\
A,B &\succeq& 0,
\ee
where $A_{\rho\sigma}^{ij}$ and $B_{\rho\sigma}^{ij}$ are respectively $N(d_1+1)\times N(d_1+1)$ and $N(d_2+1) \times N(d_2+1)$ positive semidefinite matrices.\footnote{We thank Jo\~ao Penedones for discussions of this idea.} Once again, the infinite number of (matrix) constraints is encoded in terms of finite matrices.

\section{The Conformal Bootstrap with Multiple Scalars}
\label{sec:bootstrapforscalars}

\subsection{Specializing to Scalars}
The analysis of section \ref{sec:generalsdpbootstrap} is completely general and abstract. To simplify the discussion, in this section we will focus on four-point functions of scalar operators. Here we do not assume any additional global symmetries. Let $\phi_i = (\phi_1,\phi_2,\phi_3,\ldots)$ be a collection of scalar primary fields with scaling dimensions $\Delta_i$. We consider the correlation function of four scalars,
\be
\label{eq:fourpointfunctiondecompostitionscalar}
\cG_{ijkl}(x_1,x_2,x_3,x_4) = \<\phi_i(x_1)\phi_j(x_2)\phi_k(x_3)\phi_l(x_4)\>.
\ee
The expression (\ref{eq:fourpointfunctiondecompostition}) for the four-point function simplifies because there are no representation indices $a$ in (\ref{eq:fourpointfunctiondecompostitionscalar}) and there is only one four-point structure (\ref{eq:fourptforscalars}) for four scalars. Therefore, $\cG_{ijkl}$ can be written as
\be
\cG_{ijkl}(x_1,x_2,x_3,x_4) = \frac{1}{x_{12}^{\De_i + \De_j} x_{34}^{\De_k+\De_l}} \left(\frac{x_{24}}{x_{14}}\right)^{\De_{ij}} \left(\frac{x_{14}}{x_{13}}\right)^{\De_{kl}}  G_{ijkl}(u,v).
\ee
The correlation function must be invariant under the exchange $(1,i) \leftrightarrow (3,k)$, which gives the crossing equation
\be
v^{\frac{\Delta_j+\Delta_k}{2}} G_{ijkl}(u,v) =  u^{\frac{\Delta_i + \Delta_j}{2}}G_{kjil}(v,u).
\ee
Following (\ref{eq:conformalblockexpansion}), the function $G_{ijkl}$ can be written in terms of conformal blocks $g$ as
\be
G_{ijkl}(u,v) = \sum_{\cO} \lambda_{ij\cO} \lambda_{kl\cO} g_{\Delta,\ell}^{\Delta_{ij},\Delta_{kl}}(u,v),
\ee
where compared to (\ref{eq:conformalblockexpansion}) there is no sum over different three-point structures, as only one structure appears in three-point functions containing two scalar operators. The conformal blocks also depend on the representations of the external operators $\phi_i$, but since they are scalars only the dependence on the dimensions $\Delta_i$ remains. Moreover, the conformal blocks depend only on the differences $\Delta_{ij}$ and $\Delta_{kl}$, as shown for instance in \cite{DO1,DO2} and reviewed in section \ref{sec:rational}. For real scalar external operators, the OPE coefficients $\lambda_{ij\cO}$ are real \cite{Rattazzi:2008pe}. The crossing equation is then given by
\be
\label{eq:crossyngequationintermsoff}
\sum_{\cO} \left( \lambda_{ij\cO} \lambda_{kl\cO}  v^{\frac{\Delta_j+\Delta_k}{2}} g_{\Delta,\ell}^{\Delta_{ij},\Delta_{kl}}(u,v) - \lambda_{kj\cO} \lambda_{il\cO}  u^{\frac{\Delta_i+\Delta_j}{2}} g_{\Delta,\ell}^{\Delta_{kj},\Delta_{il}}(v,u)  \right) &=& 0.
\ee
The dual form of the crossing equation is given by (\ref{eq:dualizedcrossingequationwithconformalblocks}), which for scalar operators becomes
\be
\sum_\cO \sum_{ij}\sum_{kl} \l_{ij\cO} \l_{kl\cO} \left[ \a^{ijkl}\p{ v^{\frac{\Delta_j+\Delta_k}{2}}  g_{\Delta, \ell}^{\Delta_{ij},\Delta_{kl}} } - \a^{kjil}\p{ u^{\frac{\Delta_i + \Delta_j}{2}} g_{\Delta,\ell}^{\Delta_{kj},\Delta_{il}} }  \right] &=& 0,
\ee
and can be analyzed using semidefinite programming, as explained in section \ref{sec:sdp2}.

We now introduce notation that follows more closely the notation used in the analysis of the single correlator crossing equation. Let us define functions
\be
F_{\pm,\De,\ell}^{ij,kl}(u,v)  \equiv v^{\frac{\De_k+\De_j}{2}} g_{\De,\ell}^{\De_{ij},\De_{kl}}(u,v) \pm  u^{\frac{\De_k+\De_j}{2}} g_{\De,\ell}^{\De_{ij},\De_{kl}}(v,u),
\ee
which are  respectively symmetric and antisymmetric under the exchange $u \leftrightarrow v$. They are also invariant under the simultaneous exchange of $i \leftrightarrow k$ and $j \leftrightarrow l$. In terms of these functions, the crossing equation becomes
\be\label{eq:crossingsym}
 \sum_{\cO} \left[ \l_{ij\cO} \l_{kl\cO} F^{ij,kl}_{\mp,\De,\ell}(u,v) \pm \l_{kj\cO} \l_{il\cO} F^{kj,il}_{\mp,\De,\ell}(u,v) \right]&=&0 .
\ee
If all operators are equal, the upper sign case reduces to the single correlator crossing equation (\ref{eq:sumrule}).

\subsection{Simplest System with a $\mathbb{Z}_2$ Symmetry}
Before switching to the dual form of the crossing equation, we will further simplify the system of crossing equations under consideration by assuming that the system has a $\mathbb{Z}_2$ symmetry.  Under this symmetry, all of the operators can be classified as even or odd. Let $\sigma$ and $\epsilon$ be the lowest dimension $\mathbb{Z}_2$-odd and $\mathbb{Z}_2$-even scalars, respectively. The OPE structure of these operators can be written schematically as
\bea
\sigma \times \sigma &\sim& \sum_{\cO ^+} \lambda_{\sigma \sigma \cO} \cO, \label{eq:opewithz2} \nn\\
\sigma \times \epsilon &\sim& \sum_{\cO ^-} \lambda_{\sigma \epsilon \cO} \cO, \nn\\
\epsilon \times \epsilon &\sim& \sum_{\cO ^+} \lambda_{\epsilon \epsilon \cO} \cO. 
\eea
Here $\cO^+$ runs over $\mathbb{Z}_2$-even operators of even spin and $\cO^{-}$ runs over $\mathbb{Z}_2$-odd operators of any spin. An important example of a system described by OPEs (\ref{eq:opewithz2}) is the critical Ising model, where $\sigma$ and $\epsilon$ can be thought of as the spin and energy density operators.

We now write the crossing equations (\ref{eq:crossyngequationintermsoff}) for four-point functions containing $\sigma$ and $\epsilon$. Due to the $\mathbb{Z}_2$ symmetry, some of the OPE coefficients vanish and we end up with five independent constraints:
\be
0 &=& \sum_{\cO^{+}} \l_{\s\s\cO}^2 F^{\s\s,\s\s}_{-,\De,\ell}(u,v), \nn\\
0 &=& \sum_{\cO^{+}} \l_{\e\e\cO}^2 F^{\e\e,\e\e}_{-,\De,\ell}(u,v),  \nn\\
0 &=& \sum_{\cO^{-}} \l_{\s\e\cO}^2 F^{\s\e,\s\e}_{-,\De,\ell}(u,v),  \nn\\
0 &=& \sum_{\cO^{+}}  \l_{\s\s\cO} \l_{\e\e\cO} F^{\s\s,\e\e}_{\mp,\De,\ell}(u,v) \pm \sum_{\cO^{-}} (-1)^{\ell} \l_{\s\e\cO}^2 F^{\e\s,\s\e}_{\mp,\De,\ell}(u,v),
\ee
where in the last equation we used the identity $\lambda_{\sigma\epsilon\cO} = (-1)^\ell \lambda_{\epsilon \sigma\cO}$. We can write this in vector notation as
\be
\label{eq:crossingequationwithv}
 \sum_{\cO^+} \begin{pmatrix}\l_{\s\s\cO} & \l_{\e\e\cO}\end{pmatrix} \vec{V}_{+,\De,\ell}\begin{pmatrix} \l_{\s\s\cO} \\ \l_{\e\e\cO} \end{pmatrix}+ \sum_{\cO^-} \l_{\s\e\cO}^2 \vec{V}_{-,\De,\ell} & = & 0 ,
\ee
where $\vec{V}_{-,\De,\ell}$ is a 5-vector and $\vec{V}_{+,\De,\ell}$ is a 5-vector of $2 \times 2$ matrices:
\bea
\vec{V}_{-,\De,\ell} = \begin{pmatrix} 0  \\ 0 \\ F_{-,\De,\ell}^{\s\e,\s\e}(u,v) \\ (-1)^{\ell} F_{-,\De,\ell}^{\e\s,\s\e}(u,v) \\ - (-1)^{\ell} F_{+,\De,\ell}^{\e\s,\s\e}(u,v) \end{pmatrix}, && 
\vec{V}_{+,\De,\ell} = \begin{pmatrix} \begin{pmatrix}  F^{\s\s,\s\s}_{-,\De,\ell}(u,v) & 0 \\ 0 & 0  \end{pmatrix} \\ \begin{pmatrix}  0 & 0 \\ 0 & F^{\e\e,\e\e}_{-,\De,\ell}(u,v)  \end{pmatrix}\\ \begin{pmatrix}  0 & 0 \\ 0 & 0  \end{pmatrix}  \\ \begin{pmatrix}  0 & \frac12 F^{\s\s,\e\e}_{-,\De,\ell}(u,v) \\ \frac12 F^{\s\s,\e\e}_{-,\De,\ell}(u,v) & 0 \end{pmatrix} \\ \begin{pmatrix} 0 & \frac12 F^{\s\s,\e\e}_{+,\De,\ell}(u,v) \\ \frac12 F^{\s\s,\e\e}_{+,\De,\ell}(u,v) & 0  \end{pmatrix} \end{pmatrix}.\nn\\
\eea
Let $\vec \alpha =(\alpha^1, \dots,\alpha^5)$ be a 5-vector of functionals, $\alpha^i : \cF \to \R  $. Acting on (\ref{eq:crossingequationwithv}) with this functional gives the dual form of crossing equation
\be
\sum_{\cO^+} \begin{pmatrix} \l_{\s\s\cO} & \l_{\e\e\cO}\end{pmatrix} \vec \alpha \cdot \vec{V}_{+,\De,\ell} \begin{pmatrix} \l_{\s\s\cO} \\ \l_{\e\e\cO} \end{pmatrix} + \sum_{\cO^-} \l_{\s\e\cO}^2 \vec \alpha \cdot \vec{V}_{-,\De,\ell} & =& 0,
\ee
or more explicitly
\bea
&\sum_{\cO^+} \begin{pmatrix} \l_{\s\s\cO} & \l_{\e\e\cO} \end{pmatrix}  \begin{pmatrix} \alpha^1[F_{-,\Delta,\ell}^{\s\s,\s\s}]  &  \frac{1}{2} \alpha^4[F_{-,\Delta,\ell}^{\s\s,\e\e}] + \frac{1}{2} \alpha^5[F_{+,\Delta,\ell}^{\s\s,\e\e}] \\  \frac{1}{2} \alpha^4[F_{-,\Delta,\ell}^{\s\s,\e\e}] + \frac{1}{2} \alpha^5[F_{+,\Delta,\ell}^{\s\s,\e\e}] & \alpha^2[F_{-,\Delta,\ell}^{\e\e,\e\e}] \end{pmatrix}  \begin{pmatrix} \l_{\s\s\cO} \\ \l_{\e\e\cO} \end{pmatrix} \notag\\
&+ \sum_{\cO^-} \l_{\s\e\cO}^2 \p{ \a^3[F_{-,\Delta,\ell}] + (-1)^\ell \a^4[F_{-,\Delta,\ell}^{\e\s,\s\e}] -(-1)^\ell \a^5[F_{+,\Delta,\ell}^{\e\s,\s\e}] }  = 0. \notag\\
\eea
The sum over operators in (\ref{eq:crossingequationwithv}) must contain the unit operator. For convenience we will write the unit operator contribution separately. We also assume that the operators are normalized so that $\lambda_{\s\s\mathbf{1}} = \lambda_{\e\e\mathbf{1}} = 1$. The crossing equation can then be written as
\be
\label{eq:dualcrossingequationwithv}
\begin{pmatrix} 1 & 1\end{pmatrix} \vec \alpha \cdot \vec{V}_{+,0,0} \begin{pmatrix} 1 \\ 1 \end{pmatrix} + \sum_{\cO^+} \begin{pmatrix} \l_{\s\s\cO} & \l_{\e\e\cO}\end{pmatrix} \vec \alpha \cdot \vec{V}_{+,\De,\ell} \begin{pmatrix} \l_{\s\s\cO} \\ \l_{\e\e\cO} \end{pmatrix} + \sum_{\cO^-} \l_{\s\e\cO}^2 \vec \alpha \cdot \vec{V}_{-,\De,\ell} = 0 . \nn\\
\ee

\subsection{Bounds from Semidefinite Programming}
\label{sec:convex}
Now (\ref{eq:dualcrossingequationwithv}) is in the right form for a semidefinite programming analysis as described at the end of section \ref{sec:generalsdpbootstrap}. The logic is to make an assumption about the CFT spectrum and then search for a functional $\vec \alpha$ such that
\bea
\label{eq:functionalinequalities}
\begin{pmatrix} 1 & 1\end{pmatrix} \vec \alpha \cdot \vec{V}_{+,0,0} \begin{pmatrix} 1 \\ 1 \end{pmatrix}  &>& 0 , \nn\\
\vec \alpha \cdot \vec{V}_{+,\De,\ell} &\succeq &0 ,\quad\textrm{for all $\mathbb{Z}_2$-even operators with even spin,} \nn\\
\vec \alpha \cdot \vec{V}_{-,\De,\ell} & \ge &0 ,\quad \textrm{for all $\mathbb{Z}_2$-odd operators in the spectrum.} 
\eea
If we manage to find such a functional, then crossing symmetry can not be satisfied and we conclude that the initial assumption about the spectrum is wrong. Note that (\ref{eq:functionalinequalities}) represents a sufficient condition for a functional to exclude the spectrum, but not a necessary one. In particular, it does not take into account the symmetry of the OPE coefficients, $\lambda_{\s\s\e} = \lambda_{\s\e\s}$. Using this identity we could weaken the conditions on a functional that would still violate the crossing symmetry constraints (strengthening the resulting bounds).

The space of all functionals is infinite-dimensional, so for implementation on a computer we must select a finite subspace of functionals as well as a convenient set of basis vectors. We follow the traditional approach, which has given good results in previous studies, and use linear combinations of $z$ and $\bar z$ derivatives at the point $z=\bar z=1/2$ as the basis for functionals. The functional $\a^i$ is given in that basis by
\be
\label{eq:alphaintermsofderivatives}
\alpha^i [ f ] = \sum_{m \geq n} a^i_{mn}\partial_z^m \partial_{\bar z}^n f(1/2,1/2),
\ee
where we limit the number of derivatives with the parameter $n_\text{max}$ by requiring $n+m \le 2n_\text{max}-1$.\footnote{$n_\mathrm{max}$ is equivalent to the parameters $k,K$ in \cite{Kos:2013tga, El-Showk:2014dwa}.} The sum in (\ref{eq:alphaintermsofderivatives}) only contains terms with $m\geq n$ because the functional acts on functions of $u,v$, which are symmetric under the exchange $z\leftrightarrow \bar z$. Moreover, the functional $\alpha^5$ acts on $F_+$, which is symmetric under the exchange $u\leftrightarrow v$, so only the terms with $m+n$ even are non-zero. Other components $\alpha^i$ act on functions $F_-$ antisymmetric under $u\leftrightarrow v$, so for them only the terms with $n+m$ odd are non-vanishing. The total number of independent components of the functional $\vec \alpha$ is then $N=5n_\text{max} (n_\text{max}+1)/2$. This is the dimension of the vector space in which we search for a functional $\vec \alpha$ with the properties (\ref{eq:functionalinequalities}).

To turn the constraints (\ref{eq:functionalinequalities}) into a semidefinite program suitable for numerical analysis on a computer, we employ the method described in section \ref{sec:sdp2}. We use the rational approximation of conformal blocks described explicitly in section \ref{sec:rational} to approximate the conformal blocks and their derivatives as
\be
\partial_z^m \partial_{\bar z}^n g_{\De,\ell}^{\De_{12},\De_{34}}(z=1/2,\bar z = 1/2) \aeq \chi_{\ell}(\De) p_\ell^{\De_{12},\De_{34};mn}(\Delta),
\label{eq:polynomialapproximationforblock}
\ee
where $p_\ell^{\De_{12},\De_{34};mn}(\De)$ are polynomials in $\De$ and $\chi_{\ell}(\De)$ are functions that are positive for all values of $\De$ in the CFT spectrum. Then we can similarly write
\be
\partial_z^m \partial_{\bar z}^n F_{\pm,\De,\ell}^{ij,kl}(1/2,1/2) \aeq \chi_{\ell}(\De) P_{\pm,\ell}^{ij,kl;mn}(\De),
\ee
where $P_{\pm,\ell}^{ijkl;mn}(\De)$ are polynomials related to $p_{\ell}^{\De_{ij},\De_{kl};mn}$. Using the last expression, we can rewrite the conditions (\ref{eq:functionalinequalities}) in the following form:
\bea
\begin{pmatrix} 1 & 1\end{pmatrix}Z_0(0) \begin{pmatrix} 1 \\ 1 \end{pmatrix}  &>& 0 , \nn\\
Z_{\ell}(\De) &\succeq &0 ,\quad\textrm{for all $\mathbb{Z}_2$-even operators with even spin,} \nn\\
Y_{\ell}(\De) & \ge &0 ,\quad \textrm{for all $\mathbb{Z}_2$-odd operators in the spectrum.} 
\label{eq:constraintsformixedcorrelator}
\eea
Here $Y_\ell(\De)$ are polynomials and $Z_\ell(\De)$ are matrix polynomials in $\De$ defined as
\be
Y_{\ell}(\Delta)  &\equiv& \sum_{mn} \left[ a^3_{mn} P^{\s\e,\s\e;mn}_{-,\ell}(\De) + a^4_{mn} (-1)^{\ell} P^{\e\s,\s\e,mn}_{-,\ell}(\De) - a^5_{mn} (-1)^{\ell} P^{\e\s,\s\e;mn}_{+, \ell}(\De) \right],  \notag\\
Z_{\ell}(\Delta) &\equiv&  \sum_{mn} \begin{pmatrix}a^1_{mn} P^{\s\s,\s\s;mn}_{-,\ell}(\De) & \frac12 \left( a^4_{mn} P_{-,\ell}^{\s\s,\e\e;mn}(\De) + a^5_{mn} P_{+,\ell}^{\s\s,\e\e;mn}(\De) \right)   \\ \frac12 \left( a^4_{mn} P_{-,\ell}^{\s\s,\e\e;mn}(\De) + a^5_{mn} P_{+,\ell}^{\s\s,\e\e;mn}(\De) \right) & a^2_{mn} P^{\e\e,\e\e;mn}_{-,\ell}(\De)\end{pmatrix} \notag.\\
\ee

The constraint $Y_\ell(\De)\geq 0$ can be written in terms of positive semidefinite matrices as
\be
Y_{\ell}(\Delta_{\ell}^{\text{min}}(1+x)) &=& \Tr(A_\ell Q_{d_1}(x))+x\Tr(B_\ell Q_{d_2}(x))\nn\\
&=& \sum_{\rho\s} (A_{\ell})_{\rho\s}x^{\rho+\s}+\sum_{\rho\s} (B_{\ell})_{\rho\s}x^{\rho+\s+1},
\label{eq:yequality}
\ee
where $A_\ell,B_\ell\succeq 0$ are positive semidefinite, and $d_1,d_2$ are chosen appropriately for the degree of $Y_\ell$.  Similarly, the constraint $Z_\ell(\De)\succeq 0$ can be written in terms of positive semidefinite matrices as
\be
Z_{\ell}(\Delta_{\ell}^{\text{min}}(1+x)) &=& \Tr_{\R^{d_1+1}}(C_\ell(Q_{d_1}(x)\otimes \mathbf{1}_{2\x 2}))+x\Tr_{\R^{d_2+1}}(D_\ell(Q_{d_2}(x)\otimes \mathbf{1}_{2\x2}),\nn\\
Z_{\ell}^{ij}(\Delta_{\ell}^{\text{min}}(1+x))&=& \sum_{\rho\s} (C_\ell)^{ij}_{\rho\s} x^{\rho+\s}+\sum_{\rho\s}(D_\ell)^{ij}_{\rho\s} x^{\rho+\s+1},
\label{eq:zequality}
\ee
where $C_\ell,D_\ell\succeq 0$ are positive semidefinite acting on $\R^{d_{1,2}+1}\otimes \R^2$, and $d_1,d_2$ are chosen appropriately for the degree of $Z_\ell$.  Matching terms of equal degree in $x$ on both sides, (\ref{eq:yequality}) and (\ref{eq:zequality}) become linear equations relating the variables $a_{mn}^i, A_\ell, B_\ell, C_\ell, D_\ell$.  Written in terms of these variables, our optimization problem is now in a form that can be fed to a semidefinite program solver.  We give details of our implementation in the solver \texttt{SDPA-GMP} \cite{SDPA, SDPAGMP} in appendix~\ref{app:sdpa}.

\subsection{Additional Constraints}

There are a few additional constraints on systems of multiple correlators that we have not yet incorporated into our numerical analysis.  One is that the coefficient of the stress tensor conformal block should be consistent with Ward identities.  In the OPE $\f\x\f$, Ward identities imply that the stress tensor appears with coefficient proportional to $\De_\f$.  Thus, in the conformal block decomposition for $\<\f_1\f_1\f_2\f_2\>$ in the $\f_1\f_1\to\f_2\f_2$ channel, the stress tensor block appears with a coefficient of $\l_{11T} \l_{22T} \propto \frac{\De_1\De_2}{c}$ (up to a theory-independent pre-factor that depends on our definition of the conformal block), where $c\propto\<T_{\mu\nu}T_{\rho\s}\>$ is the coefficient of the stress tensor two-point function.\footnote{$c$ is commonly referred to as the ``central charge," although it has no (known) relation to symmetry algebras in greater than $2$ spacetime dimensions.}  In the 3D Ising model, this implies that the coefficients of the stress tensor block in $\<\s\s\s\s\>$, $\<\s\s\e\e\>$, $\<\e\e\e\e\>$ must appear in the fixed ratios $\De_\s^2:\De_\s\De_\e:\De_\e^2$.

We might add this additional information to obtain stronger bounds.  However, this condition is useless without an additional assumption of a gap in the spin-2 spectrum, $\De_{T'}\geq 3+\de$ with $\de > 0$, where $T'$ is the second-lowest spin-2 operator in the theory.  In the absence of a gap, spin-2 operators with dimension $3+\varepsilon$ with $\varepsilon\ll 1$ can mimic the contribution of the stress tensor, obliterating any information about ratios of stress-tensor coefficients.  Concretely, the condition $Z_2(3+\varepsilon)\succeq 0$ in (\ref{eq:constraintsformixedcorrelator}) for all $\varepsilon>0$ also implies $Z_2(3)\succeq 0$, which is strictly stronger than necessary given the Ward identities.

In the 3D Ising model, there is indeed a large gap in the spin-2 spectrum, $\De_{T'}\gtrsim 5.5$ \cite{El-Showk:2014dwa}.  We have not  incorporated the existence of this gap and the accompanying Ward identity constraints in this work.  It will be very interesting to do so in the future.

Another constraint on the 3D Ising model is that there is precisely one operator $\e$ with dimension $\De_\e$.\footnote{We thank Slava Rychkov for explaining how to exploit this constraint.}  The condition $Z_0(\De_\e)\succeq 0$ in (\ref{eq:constraintsformixedcorrelator}) actually allows for many operators of dimension $\De_\e$ with different OPE coefficients.  In particular, it assumes only that the sum
\be
\mathop{\sum_{\dim\cO=\De_\e}}_{\mathrm{spin}\,\cO=0,\ \Z_2\textrm{-charge}\,\cO=+}\begin{pmatrix}\l_{\s\s\cO} \\ \l_{\e\e\cO}\end{pmatrix}\begin{pmatrix}\l_{\s\s\cO} & \l_{\e\e\cO}\end{pmatrix}
\label{eq:epsOPEmatrix}
\ee
is a positive semidefinite matrix (which is a consequence of unitarity).  However, because there is exactly one choice for $\cO$ above, the sum is not a generic positive semidefinite matrix: it has rank one.  Let us suppose the vector $(\l_{\s\s\e},\l_{\e\e\e})$ is proportional to $e_\th\equiv (\cos\th,\sin\th)$ for some angle $\th\in[0,\pi)$.  We can then replace the condition $Z_0(\De_\e)\succeq 0$ with the weaker condition
\be
e_\th^T Z_0(\De_\e) e_\th\geq 0.
\label{eq:rankoneconstraint}
\ee
Running our semidefinite program subject to (\ref{eq:rankoneconstraint}) will yield some allowed region $A_\th$ in the space of CFT data.  Since we do not know the actual value of $\th$ in the 3D Ising model, we must then scan over $\th$, computing the final allowed region
\be
A_* &=& \bigcup_{\th\in[0,\pi)} A_\th.
\ee
The region $A_*$ could in principle be smaller than what one gets by imposing the na\"ive condition $Z_0(\De_\e)\succeq 0$.  The idea of scanning over $\th$ to exploit the fact that (\ref{eq:epsOPEmatrix}) has rank one was initially explored in \cite{SlavaUnpublished}.  Unfortunately, performing this scan is infeasible with our current methods, given the time it takes to compute each allowed region $A_\th$.  These constraints will be important to study in the future.

Another constraint is the symmetry of three-point coefficients, in particular the relation $\l_{\s\e\s}=\l_{\s\s\e}$.  This constraint is straightforward to impose within our formalism, and we are currently exploring its consequences.

\section{Rational Representations for Conformal Blocks}
\label{sec:rational}

In order to study the semidefinite program described in the previous sections, we require systematic approximations for the derivatives of the functions $F_{\pm,\De,\ell}^{ijkl}$ in terms of positive functions times polynomials in $\De$. Such approximations directly follow from a representation for the conformal blocks as a sum over poles in $\De$. Representations of this type were first developed for 2D (Virasoro) conformal blocks by Alyosha Zamolodchikov~\cite{Zamolodchikov:1985ie,Zamolodchikov:1987}, and a generalization for global conformal blocks with identical external scalars in general spacetime dimension $D$ was developed in~\cite{Kos:2013tga}.\footnote{Other recent studies of global conformal blocks can be found in~\cite{ElShowk:2012ht,Costa:2011mg,Costa:2011dw,DO1,DO2,DO3,SimmonsDuffin:2012uy,Osborn:2012vt,Hogervorst:2013sma,Fitzpatrick:2013sya,Hogervorst:2013kva,Behan:2014dxa}, with connections to Mellin amplitudes in~\cite{Mack:2009mi,Mack:2009gy,Penedones:2010ue,Fitzpatrick:2011ia,Paulos:2011ie,Fitzpatrick:2011hu,Fitzpatrick:2011dm,Paulos:2012nu,Fitzpatrick:2012cg,Costa:2012cb}. Older work includes~\cite{Ferrara:1971vh,Ferrara:1973vz,Ferrara:1973yt,Ferrara:1974ny,Ferrara:1974nf}. Superconformal extensions have been studied in~\cite{Poland:2010wg,Berkooz:2014yda,Dolan:2001tt,Heslop:2004du,Nirschl:2004pa,Dolan:2004iy,Dolan:2004mu,Fortin:2011nq,Fitzpatrick:2014oza,Khandker:2014mpa}.} In this section we generalize the formula obtained in~\cite{Kos:2013tga} to non-identical external scalars.

Poles in the conformal block occur at special (non-unitary) dimensions $\De = \De_*$ where some descendant $P^n |\cO\>$ of the state created by the primary operator $\cO$ becomes null. This null state and all of its (also null) descendants together form a conformal sub-representation, and hence the residue of the pole is proportional to a conformal block:\footnote{At this stage it is not obvious that all such poles must be simple poles. Indeed, when the spacetime dimension $D = 2n$ is an even integer, double poles occur, while outside of even dimensions only simple poles occur. Our formulas assume $D \neq 2n$, but reproduce the correct even-$D$ conformal blocks in the limit $D \rightarrow 2n$.}
\be
g^{\De_{12},\De_{34}}_{\De,\ell} &\sim& \frac{c_\a}{\De-\De_*}g^{\De_{12},\De_{34}}_{\De_{\a}, \ell_{\a}}\textrm{\quad as \quad}\De\to\De_*.
\ee

Since poles in $\De$ determine $g^{\De_{12},\De_{34}}_{\De,\ell}$ up to a function that is analytic on the entire complex plane, we can write
\be
\label{eq:recursiong}
g^{\De_{12},\De_{34}}_{\De,\ell}(r,\eta) &=& \tilde{g}^{\De_{12},\De_{34}}_{\ell}(\De,r,\eta) + \sum_i \frac{c^{\De_{12},\De_{34}}_i}{\De-\De_i}g^{\De_{12},\De_{34}}_{\De_i+n_i,\ell_i}(r,\eta),
\ee
where $\tilde{g}^{\De_{12},\De_{34}}_{\ell}(\De,r,\eta)$ is an entire function of $\De$ and we describe the conformal cross ratios using radial coordinates~\cite{Hogervorst:2013sma}.  In Euclidean signature, where $\bar z= z^*$, these are defined by
\be
r e^{i\th} = \frac{z}{(1+\sqrt{1-z})^2},\qquad \eta=\cos\th.
\ee
The block $g^{\De_{12},\De_{34}}_{\De,\ell}(r,\eta)$ has an essential singularity of the form $r^\De$ as $\De\to\oo$.  Stripping this off, we have
\be
h^{\De_{12},\De_{34}}_{\De,\ell}(r,\eta) &\equiv& r^{-\De}g^{\De_{12},\De_{34}}_{\De,\ell}(r,\eta),\\
h^{\De_{12},\De_{34}}_{\De,\ell}(r,\eta) &=& \tilde{h}^{\De_{12},\De_{34}}_\ell(r,\eta) + \sum_i \frac{c^{\De_{12},\De_{34}}_i}{\De-\De_i}r^{n_i}h^{\De_{12},\De_{34}}_{\De_i+n_i,\ell_i}(r,\eta).
\label{eq:recursionh}
\ee
We see that the entire function $\tilde{h}_\ell^{\De_{12},\De_{34}}(r,\eta)=\lim_{\De\to\oo} h^{\De_{12},\De_{34}}_{\De,\ell}(r,\eta)$ no longer depends on $\De$, since there are no singularities as $\De\to\oo$.

Now, the function $\tilde{h}^{\De_{12},\De_{34}}_{\ell}$ can be easily computed by solving the conformal Casimir equation \cite{DO2} to leading order in $\De$, giving the result\footnote{Here we define the conformal blocks with a factor $(-1)^{\ell}$ relative to~\cite{Kos:2013tga}. In the notation of~\cite{DO3}, our blocks have a normalization coefficient of $c_{\ell} \equiv  \frac{(-1)^{\ell}}{4^{\De}}\frac{(\nu)_{\ell}}{(2\nu)_{\ell}}$.}
\be
\tilde{h}^{\De_{12}\De_{34}}_{\ell}(r,\eta) &=&  \frac{\ell!}{(2\nu)_\ell}\frac{(-1)^{\ell}C_\ell^\nu(\eta)}{(1-r^2)^\nu (1+r^2 + 2 r \eta)^{\frac12(1+\De_{12}-\De_{34})}(1+ r^2 - 2 r \eta)^{\frac12(1-\De_{12}+\De_{34})}},\nn\\
\ee
where $\nu=\frac{D-2}{2}$ and $C^\nu_\ell(\eta)$ is a Gegenbauer polynomial.
The locations of the poles $\De_i$ in (\ref{eq:recursiong}) are the same as in the case of equal external dimensions (though in that case some have vanishing coefficients) since they only depend on the representation theory of the exchanged operator. In both cases we find three sequences of poles, reproduced in table~\ref{tab:polepositions}. However, the coefficients $c^{\De_{12},\De_{34}}_i$ depend on the external dimensions and can be found by solving the conformal Casimir equation. In practice we do this order by order in the $r$ expansion, following the procedure described in~\cite{Hogervorst:2013sma}.  We compute coefficients up to high order in the $r$-expansion, guess a formula, and check the formula to even higher orders. The resulting coefficients are given by
\begin{table}
\centering
\begin{tabular}{c|c|c|cl}
$n_i$ & $\De_i$ & $\ell_i$ & $c^{\De_{12},\De_{34}}_i$\\
\cline{1-4}
$k$ & $1-\ell-k$ & $\ell+k$ & $c_1^{\De_{12},\De_{34}}(k)$  & \quad $k=1,2,\dots$\\
$2k$ & $1+\nu-k$   & $\ell$    & $c_2^{\De_{12},\De_{34}}(k)$  & \quad $k=1,2,\dots$\\ 
$k$ & $1+\ell+2\nu-k$ & $\ell-k$ & $c_3^{\De_{12},\De_{34}}(k)$ & \quad $k=1,2,\dots, \ell $
\end{tabular}
\caption{The positions of poles of $g^{\De_{12},\De_{34}}_{\De,\ell}$ in $\Delta$ and their associated data.  There are three types of poles, corresponding to the three rows in the table.  The first two types exist for all positive integer $k$, while the third type exists for positive integer $k\leq  \ell$.  The coefficients $c_1^{\De_{12},\De_{34}}(k)$, $c_2^{\De_{12},\De_{34}}(k)$, $c_3^{\De_{12},\De_{34}}(k)$ are given in (\ref{eq:poleresidues}).}
\label{tab:polepositions}
\end{table}
\be
\label{eq:poleresidues}
c_1^{\De_{12},\De_{34}}(k) &=& -\frac{4^k k (-1)^k}{(k!)^2}\frac{(\ell+2\nu)_{k} \left(\frac12 (1-k+\De_{12})\right)_{k} \left(\frac12(1-k+\De_{34})\right)_{k} }{(\ell+\nu)_{k}},\nn\\
c_2^{\De_{12},\De_{34}}(k) &=& -\frac{4^{2k} k (-1)^k}{(k!)^2} \frac{\left(\nu-k\right)_{2k}}{\left(\ell+\nu-k\right)_{2k} \left(\ell+\nu + 1 - k\right)_{2k}}\nn\\
&& \qquad \times  \left(\frac12(1-k+\ell-\De_{12} + \nu)\right)_{k} \left(\frac12(1-k+\ell+\De_{12}+\nu)\right)_{k} \nn\\
&& \qquad \times \left(\frac12(1-k+\ell-\De_{34}+\nu)\right)_{k} \left(\frac12(1-k+\ell+\De_{34}+\nu)\right)_{k},\nn\\
c_3^{\De_{12},\De_{34}}(k) &=& -\frac{4^k k (-1)^k}{(k!)^2}\frac{(\ell+1-k)_{k} \left(\frac12 (1-k+\De_{12})\right)_{k} \left(\frac12(1-k+\De_{34})\right)_{k} }{(\ell+\nu+1-k)_{k}},
\ee
where $(a)_n=\G(a+n)/\G(a)$ denotes the Pochhammer symbol.
It should be possible to analytically derive these coefficients using conformal representation theory.  A derivation could shed light on their generalization to other conformal blocks (e.g., for operators with spin) and superconformal blocks.

Using the recursion relation (\ref{eq:recursionh}), it is straightforward to compute derivatives of the conformal blocks around the crossing-symmetric point $r=r_*=3-2\sqrt 2\aeq 0.17$, $\eta=1$.\footnote{The computation for $\De_{12}=\De_{34}=0$ is described explicitly in \cite{El-Showk:2014dwa}.  Once we fix $\De_{12},\De_{34}$ the computation here is essentially the same.  We must compute different tables of derivatives for each numerical value of $\De_{12},\De_{34}$.}  These have the form
\be
\ptl_z^m \ptl_{\bar z}^n g_{\De,\ell}^{\De_{12},\De_{34}}(r_*,1) &=& r_*^\De\p{q_\ell^{\De_{12},\De_{34};mn}(\De)+\sum_i \frac{a_{\ell i}^{\De_{12},\De_{34};mn}}{\De-\De_i}},
\label{eq:partialfraction}
\ee
where $q_\ell^{\De_{12},\De_{34};mn}(\De)$ is a polynomial in $\De$ and $a_{\ell i}^{\De_{12},\De_{34};mn}$ are numerical coefficients.  Poles corresponding to larger values of $n_i$ in table~\ref{tab:polepositions} are suppressed by higher powers of $r_*\aeq 0.17$.  Thus, we can get a good approximation by truncating to a finite number of poles $n_i\leq \nu_\mathrm{max}$, with the result
\be
\ptl_z^m \ptl_{\bar z}^n g_{\De,\ell}^{\De_{12},\De_{34}}(r_*,1) &\aeq& \frac{r_*^\De}{\prod_i(\De-\De_i)} p^{\De_{12},\De_{34};mn}_\ell(\De),
\ee
where $p^{\De_{12},\De_{34};mn}_\ell(\De)$ is a polynomial obtained by combining the poles in the partial fraction expansion~(\ref{eq:partialfraction}).\footnote{The accuracy can be improved further by applying the technique described in appendix A of \cite{Kos:2013tga}.}  This has the form (\ref{eq:polynomialapproximationforblock}), with $\chi_\ell(\De)=r_*^\De \prod_i (\De-\De_i)^{-1}$.  The accuracy of this approximation depends on $\nu_\mathrm{max}$.    Bounds involving more derivatives (higher $n_\mathrm{max}$) require more precise expressions for the blocks, and consequently higher $\nu_\mathrm{max}$.  We have verified that the bounds computed in this work are essentially unchanged if $\nu_\mathrm{max}$ is increased.

\section{Results and 3D Ising Interpretation}
\label{sec:results}
\subsection{General Bound on $\Delta_\epsilon$}
Previous numerical bootstrap studies considered only a single four-point function $\<\s\s\s\s\>$. One of the results of the single correlator bootstrap is a rigorous upper bound on the dimension of the lowest-dimension scalar $\e$ appearing in the $\s\x\s$ OPE. The bound on $\De_\e$ as a function of $\De_\s$ was obtained for $D=3$ in \cite{ElShowk:2012ht}; we reproduce it here and plot it in several figures, as explained below.

In the case of multiple correlators, both $\s$ and $\e$ appear as external operators, resulting in the system of equations (\ref{eq:crossingequationwithv}). It is clear that the bound resulting from this system will be at least as strong as the single-correlator bound because we can set $\alpha^{2,3,4,5}=0$ in (\ref{eq:dualcrossingequationwithv}) to reduce it to the single-correlator problem. A priori, it is possible for the complete system (\ref{eq:crossingequationwithv}) to give an even stronger bound on $\De_\e$. However, after proceeding with the computations described in section~\ref{sec:convex}, assuming nothing else about the spectrum except that $\s$ and $\e$ are respectively the lowest dimension $\mathbb{Z}_2$-odd and -even scalars,\footnote{Additionally, we always include the usual unitarity bounds on operator dimensions.} we find that the multi-correlator and single-correlator bounds agree (at $n_\mathrm{max}=6$).

While it is not obvious what the allowed region should be, it is clear that $\Z_2$-symmetry combined with multiple correlator constraints should not fix the dimensions $(\De_\s,\De_\e)$ without additional assumptions.  For example, the $O(N)$ vector models admit a $\Z_2$ symmetry, so they should lie in the allowed region, but below the 3D Ising model.\footnote{To apply our bounds to the $O(N)$ models, we may take $\s=\f_1$ and $\e=S_{11}$, where $\f_i$ is a vector under $O(N)$ and $S_{ij}$ is the lowest-dimension traceless symmetric tensor of $O(N)$ in $\f_i\x\f_j$.  Of course, much stronger bounds can be obtained by using the full information of $O(N)$ symmetry, as in \cite{Kos:2013tga}.}

\subsection{Bound on $\Delta_{\sigma'}$}
A new feature of the multiple correlator bootstrap for the Ising model is the access to the $\mathbb{Z}_2$-odd spectrum. In particular, this allows us to place an upper bound on the dimension of the second $\mathbb{Z}_2$-odd scalar $\s '$. We assume that all $\mathbb{Z}_2$-odd scalars (other than $\s,\s'$) have dimensions greater than $\De_{\s'}$ and try to find a contradiction with the crossing equation (\ref{eq:crossingequationwithv}). In the notation of section~\ref{sec:convex}, we must find a functional satisfying the constraints:
\bea
\begin{pmatrix} 1 & 1\end{pmatrix}Z_0(0) \begin{pmatrix} 1 \\ 1 \end{pmatrix}  &>& 0 , \nn\\
Z_{0}(\De) &\succeq &0,\quad \forall \De \ge \De_\e , \nn\\
 Z_{\ell}(\De) &\succeq& 0, \quad \forall \De \ge \De^{\text{min}}_\ell  , \quad\ell =2,4,\dots , \nn\\
Y_{0}(\De_\s) & \ge &0, \nn\\
\qquad Y_{0}(\De) &\ge& 0, \quad \forall \De\ge\De_{\s'}  ,\nn\\
\qquad Y_{\ell}(\De) &\ge& 0, \quad \forall \De\ge\De^{\text{min}}_\ell  , \quad\ell =1,2,\dots ,
\label{eq:sigmaprimeconstraints}
\eea
where $\De^{\text{min}}_\ell$ is the unitarity bound for a spin-$\ell$ operator. For given $\De_\s$ and $\De_\e$, we find the minimal value of $\De_{\s'}$ for which the spectrum is excluded. This value, depending on both $\De_{\s}$ and $\De_{\e}$, is an upper bound on the dimension of $\s'$. 

Instead of making a three-dimensional plot of $\De_{\s'}$ vs. $(\De_\s, \De_\e)$, we choose a curve in the $(\De_\s, \De_\e)$ plane and plot the $\De_{\s'}$ bound along it.  In particular, we choose $(\De_\s,\De_\e)$ to lie on the (single-correlator) upper bound on $\De_\e$ computed at $n_\mathrm{max}=10$ --- the black dotted line in figures~\ref{fig:effectOfEpsPrimeGap} and \ref{fig:effectOfEpsPrimeGapZoomed}.  This choice of curve is somewhat arbitrary.  An advantage is that it should pass near the 3D Ising point. Our primary goal is to get a general picture of the constraints on $\De_{\s'}$.

\begin{figure}[t!]
\begin{center}
\begin{psfrags}
\def\PFGstripminus-#1{#1}%
\def\PFGshift(#1,#2)#3{\raisebox{#2}[\height][\depth]{\hbox{%
  \ifdim#1<0pt\kern#1 #3\kern\PFGstripminus#1\else\kern#1 #3\kern-#1\fi}}}%
\providecommand{\PFGstyle}{}%
\psfrag{title}[cc][cc]{\PFGstyle $\text{upper bound on $\De_{\s'}$ ($n_\mathrm{max}=6$)}$}
\psfrag{xLabel}[cl][cl]{\PFGstyle $\text{$\De_\s$}$}
\psfrag{yLabel}[bc][bc]{\PFGstyle $\text{$\De_{\s'}$}$}
\psfrag{x05}[tc][tc]{\PFGstyle $0.5$}
\psfrag{x0505}[tc][tc]{\PFGstyle $0.505$}
\psfrag{x051}[tc][tc]{\PFGstyle $0.51$}
\psfrag{x0515}[tc][tc]{\PFGstyle $0.515$}
\psfrag{x052}[tc][tc]{\PFGstyle $0.52$}
\psfrag{x0525}[tc][tc]{\PFGstyle $0.525$}
\psfrag{x053}[tc][tc]{\PFGstyle $0.53$}
\psfrag{x0535}[tc][tc]{\PFGstyle $0.535$}
\psfrag{y1}[cr][cr]{\PFGstyle $1$}
\psfrag{y2}[cr][cr]{\PFGstyle $2$}
\psfrag{y3}[cr][cr]{\PFGstyle $3$}
\psfrag{y4}[cr][cr]{\PFGstyle $4$}
\psfrag{y5}[cr][cr]{\PFGstyle $5$}
\psfrag{y6}[cr][cr]{\PFGstyle $6$}
\psfrag{y7}[cr][cr]{\PFGstyle $7$}
\includegraphics[width=0.9\textwidth]{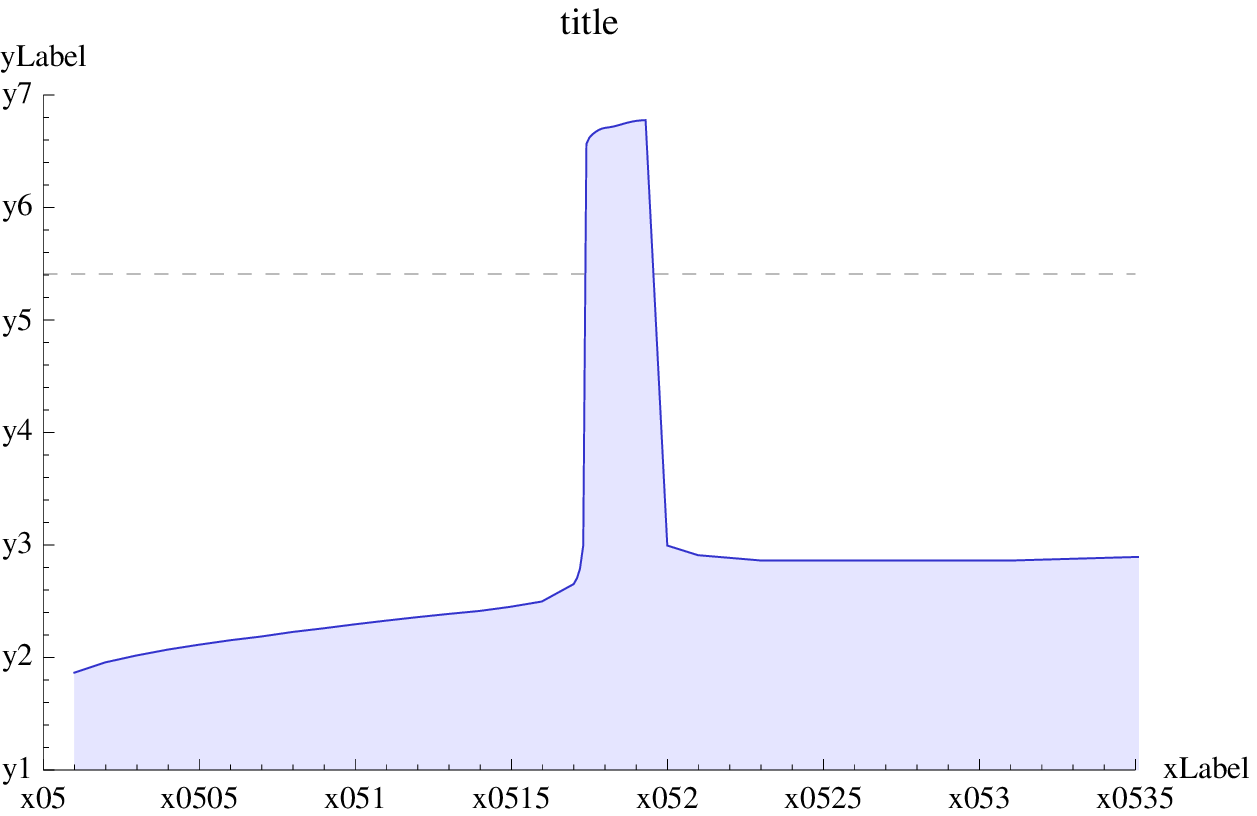}
\end{psfrags}
\caption{An upper bound on $\De_{\s'}$, where $(\De_\s, \De_\e)$ are constrained to lie on the $n_\mathrm{max}=10$ single correlator bound (black dotted line in figures~\ref{fig:effectOfEpsPrimeGap} and \ref{fig:effectOfEpsPrimeGapZoomed}).  The sharp spike in the bound occurs when the values of $\De_\s,\De_\e$ lie within the allowed region in figure~\ref{fig:effectOfEpsPrimeGap}, with no assumption about $\Z_2$-even gaps (medium-blue shaded region).  Only when $\De_\s,\De_\e$ take these values, is it possible to have $\De_{\s'}\geq 3$, and indeed the upper bound on $\De_{\s'}$ becomes extremely weak.  This bound is computed at $n_\mathrm{max}=6$, $\nu_\mathrm{max}=8$.  We expect that as $n_\mathrm{max}$ increases, it becomes more sharply peaked, while the top moves down to the correct value of $\De_{\s'}$ in the 3D Ising model. At $n_\mathrm{max}=10,\nu_\mathrm{max}=14$ we have computed the stronger bound $\De_{\s'} \leq 5.41(1)$ (the dashed line) at the points $(\De_\s,\De_\e)=(0.5181, 1.41206)$, $(0.51815, 1.41267)$, $(0.5182, 1.41312)$.}
\label{fig:sigmaPrimeBoundNmax6}
\end{center}
\end{figure}

The bound on $\De_{\s'}$ as a function of $\De_\s$ (with $\De_\e$ following the $n_\mathrm{max}=10$ single correlator bound) is shown in figure~\ref{fig:sigmaPrimeBoundNmax6} at $n_\text{max}=6$, corresponding to a linear functional with $N=105$ components.\footnote{For lower values of $n_\text{max}$ the bound only exists for $\De_\s<0.51$.}  It has an almost rectangular peak centered near the Ising value of $\De_\s$. The sides of the peak are close to vertical and the top is relatively flat.  The width of the peak is partially an artifact of our choice of $(\De_\s,\De_\e)$ curve, as will be clear in the next subsection.

This peak provides another example of interesting behavior in bootstrap bounds near the Ising point. Away from the Ising point, for $\De_\s<0.517$ or $\De_\s>0.52$, the bound on $\De_{\s'}$ is quite strong, implying $\De_{\s'}<3$ for the range of $\De_\s$ plotted in figure~\ref{fig:sigmaPrimeBoundNmax6}.  Just at the Ising point, the bound $\De_{\s'} \lesssim 6.5$ is relatively weak compared to the expected value of $\De_{\s'} = 3 + \omega_A$ in the 3D Ising CFT, which has been estimated to be $\De_{\s'} \gtrsim 4.5$ by resumming the $\epsilon$-expansion~\cite{Pelissetto:2000ek}, $\De_{\s'} \approx 5.4$ using the scaling field approach~\cite{Newman:1984zz}, and $\De_{\s'} \approx 4.7 - 5.2$ using exact RGE methods~\cite{Litim:2003kf}.

Enlarging the search space of functionals by icreasing $n_\text{max}$, the peak gets narrower and smaller.  At $n_\mathrm{max}=10$, we computed a bound on $\De_{\s'}$ to precision $0.01$ at the three points $(\De_\s,\De_\e)=(0.5181, 1.41206)$, $(0.51815, 1.41267)$, $(0.5182, 1.41312)$ near the expected values in the 3D Ising model.  At each of these points we find $\De_{\s'} \leq 5.41(1)$, closer to the estimates of other methods, but perhaps still somewhat larger.  The difference in the $n_\mathrm{max}=6,10$ results indicates that this bound has yet to converge to its optimal value and could likely be improved with further numerical work.

\subsection{Bounds on $(\De_\s, \De_\e)$ with Gaps in the Operator Spectrum}
The $\De_{\s'}$ bound in figure~\ref{fig:sigmaPrimeBoundNmax6} indicates that away from the Ising point the spectrum must contain a $\mathbb{Z}_2$-odd scalar of dimension $\De_{\s'}<3$. Conversely, imposing $\De_{\s'}\ge3$ will  exclude values of $\De_\s$ sufficiently far from the Ising model. The assumption $\De_{\s'}\ge3$ has physical meaning: it implies that there is only one relevant $\mathbb{Z}_2$-odd operator, $\s$.  This assumption is known to hold for the critical Ising model, where the only relevant operators are $\s$ and $\e$. Using this as input, we can additionally assume a gap $\De_\e \ge 3$ in the $\mathbb{Z}_2$-even spectrum to obtain even stronger bounds on scaling dimensions of the Ising model.

\begin{figure}[t!]
\begin{center}
\begin{psfrags}
\def\PFGstripminus-#1{#1}%
\def\PFGshift(#1,#2)#3{\raisebox{#2}[\height][\depth]{\hbox{%
  \ifdim#1<0pt\kern#1 #3\kern\PFGstripminus#1\else\kern#1 #3\kern-#1\fi}}}%
\providecommand{\PFGstyle}{}%
\psfrag{title}[cc][cc]{\PFGstyle $\text{allowed region with $\De_{\s'}\geq 3$ ($n_\mathrm{max}=6$)}$}
\psfrag{xLabel}[cl][cl]{\PFGstyle $\text{$\De_\s$}$}
\psfrag{yLabel}[bc][bc]{\PFGstyle $\text{$\De_\e$}$}
\psfrag{x05}[tc][tc]{\PFGstyle $0.5$}
\psfrag{x051}[tc][tc]{\PFGstyle $0.51$}
\psfrag{x052}[tc][tc]{\PFGstyle $0.52$}
\psfrag{x053}[tc][tc]{\PFGstyle $0.53$}
\psfrag{x054}[tc][tc]{\PFGstyle $0.54$}
\psfrag{x055}[tc][tc]{\PFGstyle $0.55$}
\psfrag{x056}[tc][tc]{\PFGstyle $0.56$}
\psfrag{x057}[tc][tc]{\PFGstyle $0.57$}
\psfrag{x058}[tc][tc]{\PFGstyle $0.58$}
\psfrag{x059}[tc][tc]{\PFGstyle $0.59$}
\psfrag{x06}[tc][tc]{\PFGstyle $0.6$}
\psfrag{y1}[cr][cr]{\PFGstyle $1$}
\psfrag{y11}[cr][cr]{\PFGstyle $1.1$}
\psfrag{y12}[cr][cr]{\PFGstyle $1.2$}
\psfrag{y13}[cr][cr]{\PFGstyle $1.3$}
\psfrag{y14}[cr][cr]{\PFGstyle $1.4$}
\psfrag{y15}[cr][cr]{\PFGstyle $1.5$}
\psfrag{y16}[cr][cr]{\PFGstyle $1.6$}
\psfrag{y17}[cr][cr]{\PFGstyle $1.7$}
\includegraphics[width=0.9\textwidth]{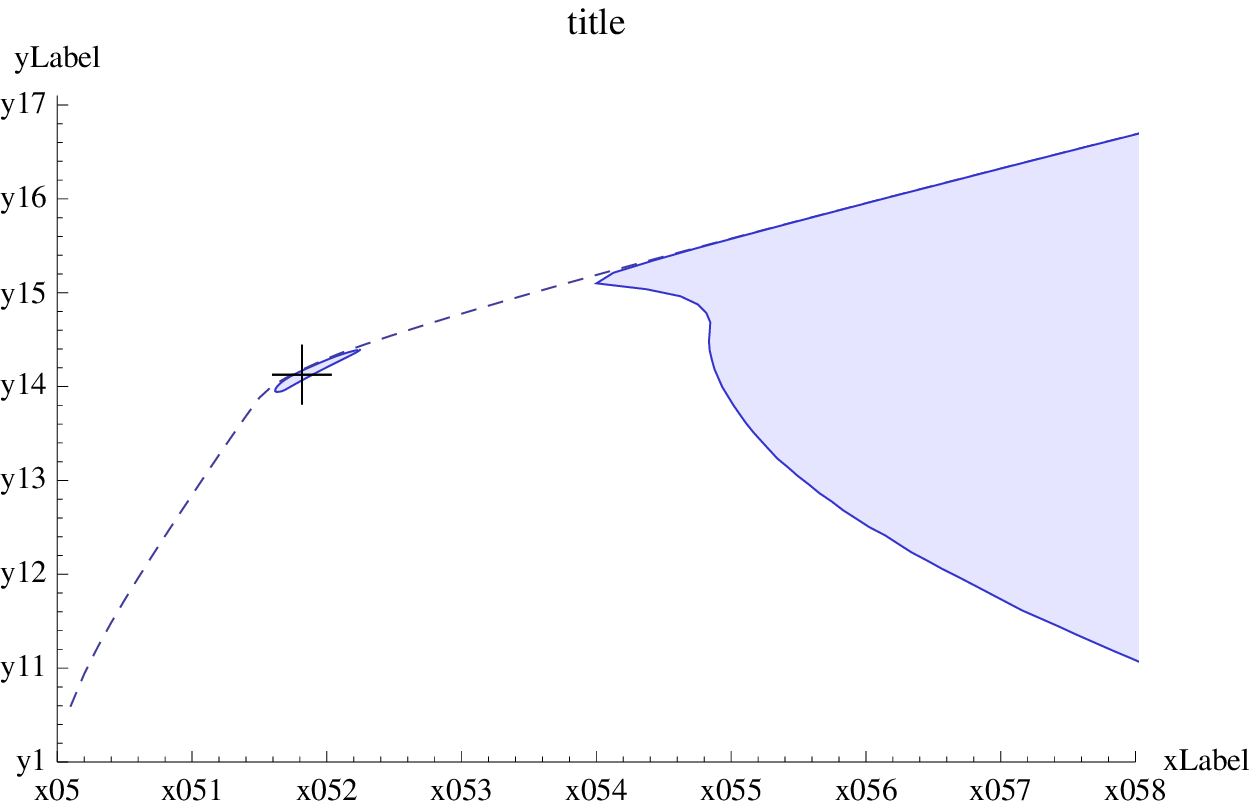}
\end{psfrags}
\caption{Allowed region of $(\De_\s,\De_\e)$ in a $\Z_2$-symmetric CFT$_3$ where $\De_{\s'}\geq 3$ (only one $\Z_2$-odd scalar is relevant). This bound uses crossing symmetry and unitarity for $\<\s\s\s\s\>$, $\<\s\s\e\e\>$, and $\<\e\e\e\e\>$, with $n_\mathrm{max}=6$ (105-dimensional functional), $\nu_\mathrm{max}=8$. The 3D Ising point is indicated with black crosshairs.  The gap in the $\Z_2$-odd sector is responsible for creating a small closed region around the Ising point.}
\label{fig:nmax6MulticorrelatorRegionPlot}
\end{center}
\end{figure}

With the assumption of a gap in the $\mathbb{Z}_2$-odd spectrum, we find a strong constraint on the values of $\De_\s$ and $\De_\e$. The allowed region in the $(\De_\s,\De_\e)$ plane is shown shaded light blue in figure~\ref{fig:nmax6MulticorrelatorRegionPlot} for $n_{\text{max}}=6$. As expected from the $\De_{\s'}$ bound, it consists of a small closed region around the Ising point and another big region at $\De_\s \gtrsim 0.54$. The dashed line is the single correlator bound. Note that since we expect $\De_{\s'}\approx 4.5$ in the Ising model, we could have assumed a bigger gap in the $\mathbb{Z}_2$-odd sector to get an even smaller allowed region in the $(\De_\s, \De_\e)$ plane. However, due to the fact that the sides of the peak in the $\De_{\s'}$ bound in figure~\ref{fig:sigmaPrimeBoundNmax6} are almost vertical, we expect that the allowed region around the Ising point is not significantly affected by the exact value of the $\mathbb{Z}_2$-odd gap, as long as it is greater than 3.

\begin{figure}[t!]
\begin{center}
\begin{psfrags}
\def\PFGstripminus-#1{#1}%
\def\PFGshift(#1,#2)#3{\raisebox{#2}[\height][\depth]{\hbox{%
  \ifdim#1<0pt\kern#1 #3\kern\PFGstripminus#1\else\kern#1 #3\kern-#1\fi}}}%
\providecommand{\PFGstyle}{}%
\psfrag{title}[cc][cc]{\PFGstyle $\text{allowed region with various gaps in $\De_{\e'},\De_{\s'}$ ($n_\mathrm{max}=6$)}$}
\psfrag{xLabel}[cl][cl]{\PFGstyle $\text{$\De_\s$}$}
\psfrag{yLabel}[bc][bc]{\PFGstyle $\text{$\De_\e$}$}
\psfrag{x05}[tc][tc]{\PFGstyle $0.5$}
\psfrag{x0505}[tc][tc]{\PFGstyle $0.505$}
\psfrag{x051}[tc][tc]{\PFGstyle $0.51$}
\psfrag{x0515}[tc][tc]{\PFGstyle $0.515$}
\psfrag{x052}[tc][tc]{\PFGstyle $0.52$}
\psfrag{x0525}[tc][tc]{\PFGstyle $0.525$}
\psfrag{x053}[tc][tc]{\PFGstyle $0.53$}
\psfrag{y1}[cr][cr]{\PFGstyle $1$}
\psfrag{y11}[cr][cr]{\PFGstyle $1.1$}
\psfrag{y12}[cr][cr]{\PFGstyle $1.2$}
\psfrag{y13}[cr][cr]{\PFGstyle $1.3$}
\psfrag{y14}[cr][cr]{\PFGstyle $1.4$}
\psfrag{y15}[cr][cr]{\PFGstyle $1.5$}
\psfrag{y16}[cr][cr]{\PFGstyle $1.6$}
\psfrag{y17}[cr][cr]{\PFGstyle $1.7$}
\psfrag{SC}[cc][cc]{\PFGstyle $\text{Single Correlator}$}
\psfrag{DE3}[cc][cc]{\PFGstyle $\text{$\De_{\e'}\geq 3$}$}
\includegraphics[width=0.9\textwidth]{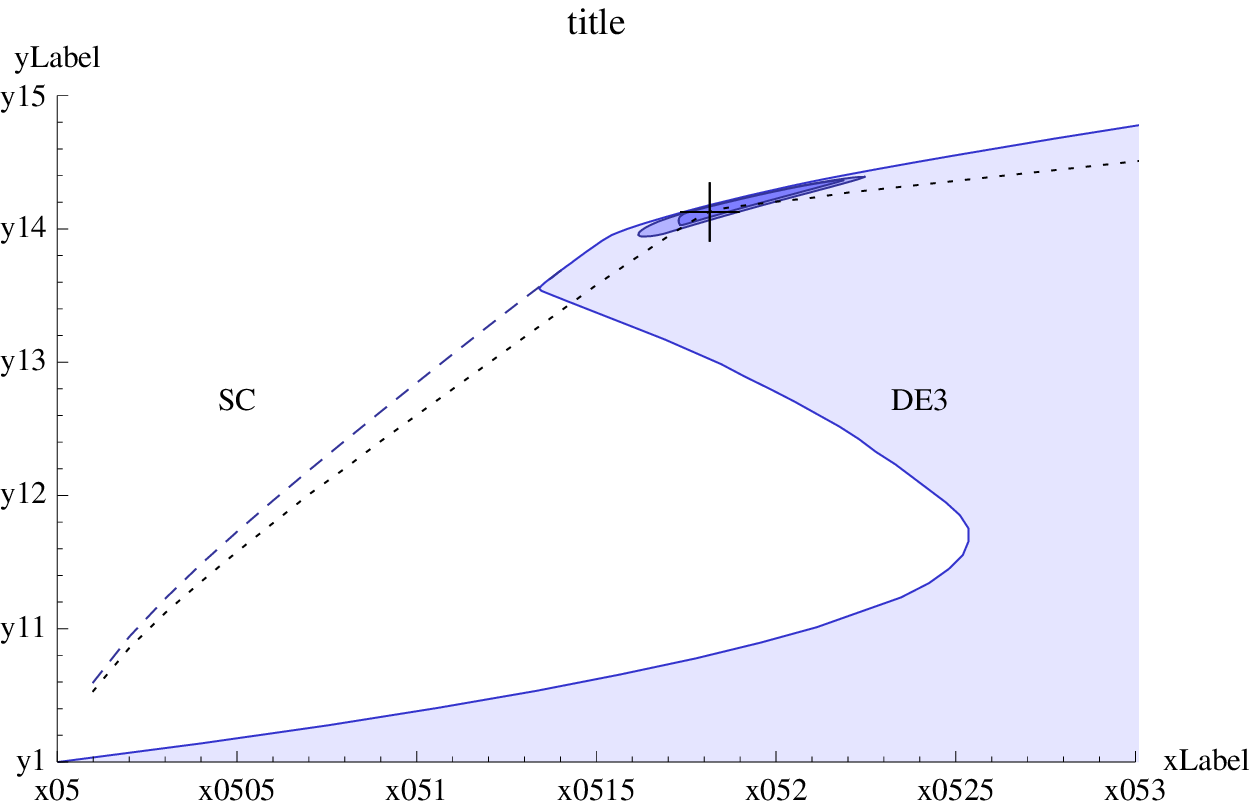}
\end{psfrags}
\caption{Allowed regions in a $\Z_2$-symmetric CFT${}_3$, assuming various gaps in the scalar spectrum.  The dashed line is an upper bound on $\De_\e$ using crossing symmetry and unitarity of $\<\s\s\s\s\>$, with no assumptions about gaps, at $n_\mathrm{max}=6$.  The black dotted line is the same bound with $n_\mathrm{max}=10$.  The light blue shaded region assumes a gap $\De_{\e'}\geq 3$ in the $\Z_2$-even sector.  The medium blue shaded region assumes a gap $\De_{\s'}\geq 3$ in the $\Z_2$-odd sector, and uses crossing symmetry for the system of correlators $\<\s\s\s\s\>,\<\s\s\e\e\>,\<\e\e\e\e\>$ (same as figure~\ref{fig:nmax6MulticorrelatorRegionPlot}).  The dark blue region assumes both $\De_{\s'},\De_{\e'}\geq 3$, and uses the system of multiple correlators. All bounds other than the black dotted line are computed with $n_\mathrm{max}=6$, $\nu_\mathrm{max}=8$ (21 components for single correlator bounds, 105 components for multiple correlator bounds).  The 3D Ising point is indicated with black crosshairs.}
\label{fig:effectOfEpsPrimeGap}
\end{center}
\end{figure}

\begin{figure}[t!]
\begin{center}
\begin{psfrags}
\def\PFGstripminus-#1{#1}%
\def\PFGshift(#1,#2)#3{\raisebox{#2}[\height][\depth]{\hbox{%
  \ifdim#1<0pt\kern#1 #3\kern\PFGstripminus#1\else\kern#1 #3\kern-#1\fi}}}%
\providecommand{\PFGstyle}{}%
\psfrag{title}[cc][cc]{\PFGstyle $\text{allowed region with various gaps in $\De_{\e'},\De_{\s'}$, zoom ($n_\mathrm{max}=6$)}$}
\psfrag{xLabel}[cl][cl]{\PFGstyle $\text{$\De_\s$}$}
\psfrag{yLabel}[bc][bc]{\PFGstyle $\text{$\De_\e$}$}
\psfrag{x0516}[tc][tc]{\PFGstyle $0.516$}
\psfrag{x0517}[tc][tc]{\PFGstyle $0.517$}
\psfrag{x0518}[tc][tc]{\PFGstyle $0.518$}
\psfrag{x0519}[tc][tc]{\PFGstyle $0.519$}
\psfrag{x052}[tc][tc]{\PFGstyle $0.52$}
\psfrag{x0521}[tc][tc]{\PFGstyle $0.521$}
\psfrag{x0522}[tc][tc]{\PFGstyle $0.522$}
\psfrag{x0523}[tc][tc]{\PFGstyle $0.523$}
\psfrag{y139}[cr][cr]{\PFGstyle $1.39$}
\psfrag{y14}[cr][cr]{\PFGstyle $1.4$}
\psfrag{y141}[cr][cr]{\PFGstyle $1.41$}
\psfrag{y142}[cr][cr]{\PFGstyle $1.42$}
\psfrag{y143}[cr][cr]{\PFGstyle $1.43$}
\psfrag{y144}[cr][cr]{\PFGstyle $1.44$}
\includegraphics[width=0.9\textwidth]{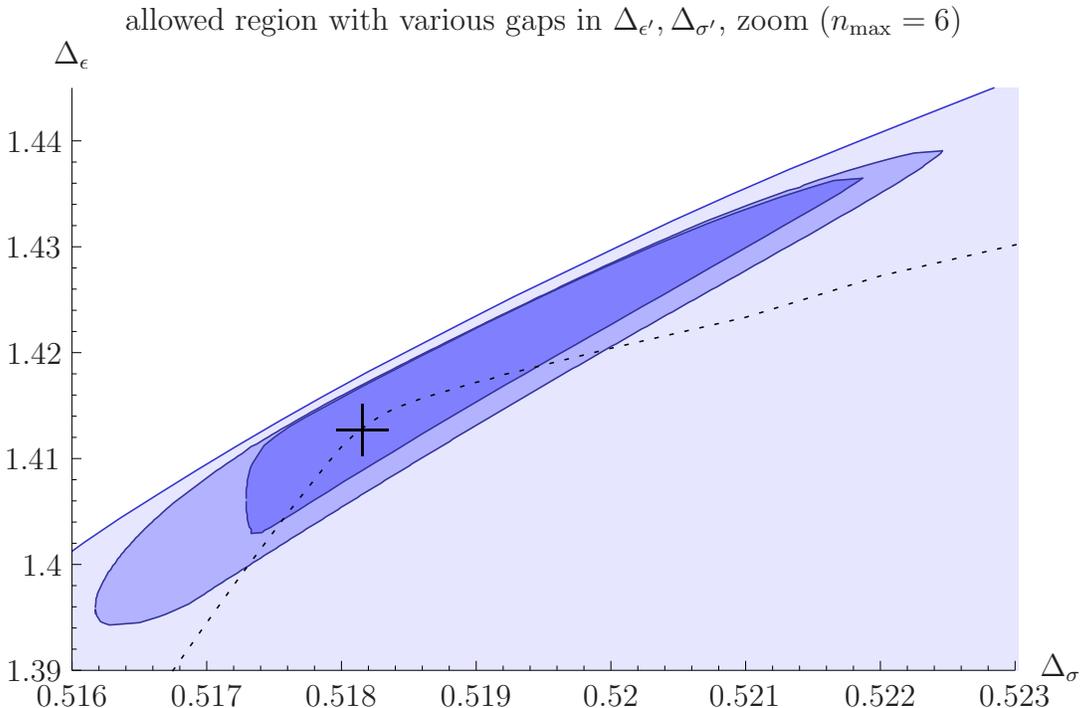}
\end{psfrags}
\caption{Zoom in on the region of the 3D Ising point in figure~\ref{fig:effectOfEpsPrimeGap}.}
\label{fig:effectOfEpsPrimeGapZoomed}
\end{center}
\end{figure}

The effect of the gap in the $\mathbb{Z}_2$-even sector is shown in figure~\ref{fig:effectOfEpsPrimeGap}. Without the $\mathbb{Z}_2$-odd gap, we get exactly the same bound that was obtained for the single correlator bootstrap in \cite{ElShowk:2012ht}. The allowed region for that case is shaded light blue in figure~\ref{fig:effectOfEpsPrimeGap}. Assuming gaps in both the $\mathbb{Z}_2$-odd and -even parts of the spectrum, we find the allowed region around the Ising point (shaded dark blue) to be of similar shape, but somewhat smaller size than the allowed region when assuming only the $\mathbb{Z}_2$-odd gap (shaded medium blue). A zoomed version of this region is shown in figure~\ref{fig:effectOfEpsPrimeGapZoomed}.

\begin{figure}[t!]
\begin{center}
\begin{psfrags}
\def\PFGstripminus-#1{#1}%
\def\PFGshift(#1,#2)#3{\raisebox{#2}[\height][\depth]{\hbox{%
  \ifdim#1<0pt\kern#1 #3\kern\PFGstripminus#1\else\kern#1 #3\kern-#1\fi}}}%
\providecommand{\PFGstyle}{}%
\psfrag{title}[cc][cc]{\PFGstyle $\text{allowed region with $\De_{\e'},\De_{\s'}\geq 3$ ($n_\mathrm{max}=10$)}$}
\psfrag{xLabel}[cl][cl]{\PFGstyle $\text{$\De_\s$}$}
\psfrag{yLabel}[bc][bc]{\PFGstyle $\text{$\De_\e$}$}
\psfrag{x05179}[tc][tc]{\PFGstyle $0.5179$}
\psfrag{x0518}[tc][tc]{\PFGstyle $0.518$}
\psfrag{x05181}[tc][tc]{\PFGstyle $0.5181$}
\psfrag{x05182}[tc][tc]{\PFGstyle $0.5182$}
\psfrag{x05183}[tc][tc]{\PFGstyle $0.5183$}
\psfrag{x05184}[tc][tc]{\PFGstyle $0.5184$}
\psfrag{x05185}[tc][tc]{\PFGstyle $0.5185$}
\psfrag{x05186}[tc][tc]{\PFGstyle $0.5186$}
\psfrag{y141}[cr][cr]{\PFGstyle $1.41$}
\psfrag{y1411}[cr][cr]{\PFGstyle $1.411$}
\psfrag{y1412}[cr][cr]{\PFGstyle $1.412$}
\psfrag{y1413}[cr][cr]{\PFGstyle $1.413$}
\psfrag{y1414}[cr][cr]{\PFGstyle $1.414$}
\psfrag{y1415}[cr][cr]{\PFGstyle $1.415$}
\includegraphics[width=0.9\textwidth]{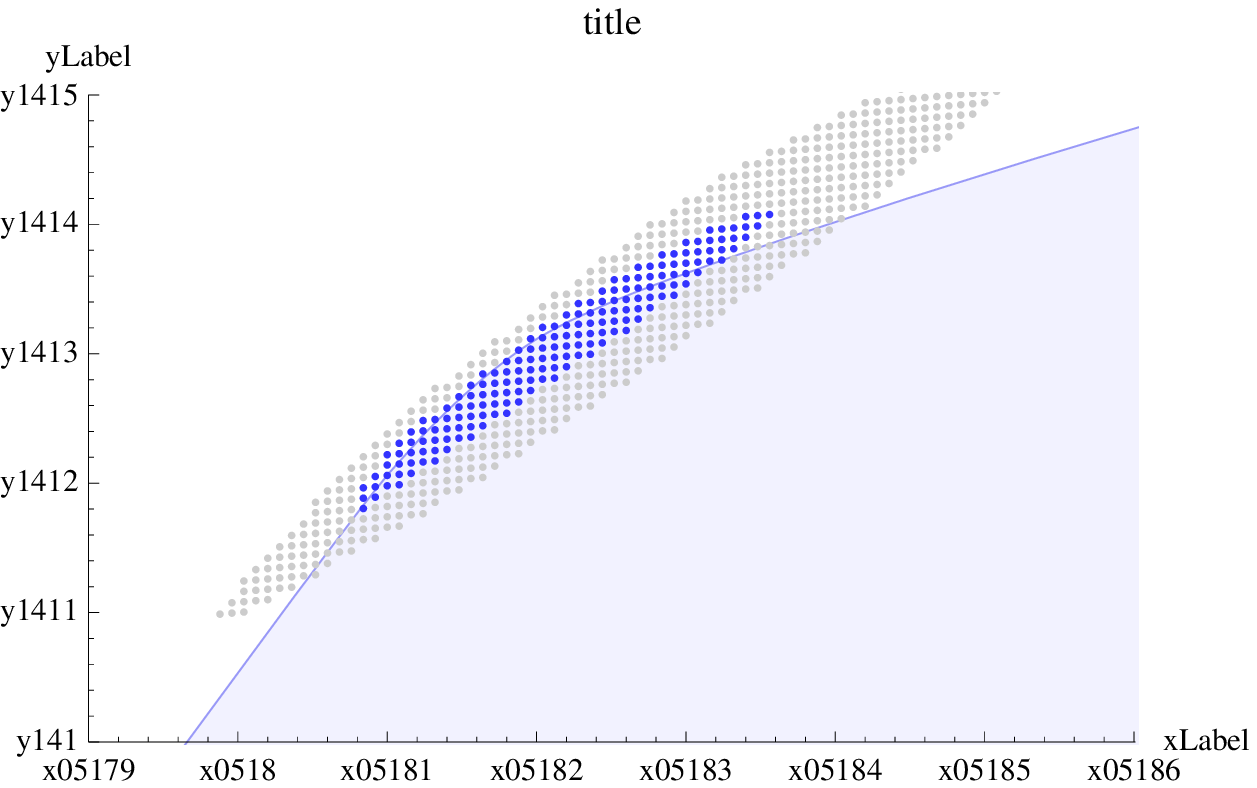}
\end{psfrags}
\caption{Allowed and disallowed $(\De_\s,\De_\e)$ points in a $\Z_2$-symmetric CFT${}_3$ with only one relevant $\Z_2$-odd and $\Z_2$-even scalar, using the constraints of crossing symmetry and unitarity for $\<\s\s\s\s\>$, $\<\s\s\e\e\>$, $\<\e\e\e\e\>$ at $n_\mathrm{max}=10$ (275 components), $\nu_\mathrm{max}=14$.  The light grey points are ruled out, while the dark blue points are allowed.  The light blue shaded region shows the region allowed by crossing symmetry and unitarity of the single correlator $\<\s\s\s\s\>$ at $n_\mathrm{max}=14$, computed in \cite{El-Showk:2014dwa}.  The final allowed region is the intersection of this shaded region with the region indicated by the dark blue points (see figure~\ref{fig:allowedRegionNmax10Summary}).\protect\footnotemark}
\label{fig:allowedRegionNmax10}
\end{center}
\end{figure}

\begin{figure}[t!]
\begin{center}
\begin{psfrags}
\def\PFGstripminus-#1{#1}%
\def\PFGshift(#1,#2)#3{\raisebox{#2}[\height][\depth]{\hbox{%
  \ifdim#1<0pt\kern#1 #3\kern\PFGstripminus#1\else\kern#1 #3\kern-#1\fi}}}%
\providecommand{\PFGstyle}{}%
\psfrag{title}[cc][cc]{\PFGstyle $\text{allowed region with $\De_{\e'},\De_{\s'}\geq 3$ ($n_\mathrm{max}=10$)}$}
\psfrag{xLabel}[cl][cl]{\PFGstyle $\text{$\De_\s$}$}
\psfrag{yLabel}[bc][bc]{\PFGstyle $\text{$\De_\e$}$}
\psfrag{x051805}[tc][tc]{\PFGstyle {\small $0.51805$}}
\psfrag{x05181}[tc][tc]{\PFGstyle {\small $0.5181$}}
\psfrag{x051815}[tc][tc]{\PFGstyle {\small $0.51815$}}
\psfrag{x05182}[tc][tc]{\PFGstyle {\small $0.5182$}}
\psfrag{x051825}[tc][tc]{\PFGstyle {\small $0.51825$}}
\psfrag{x05183}[tc][tc]{\PFGstyle {\small $0.5183$}}
\psfrag{x051835}[tc][tc]{\PFGstyle {\small $0.51835$}}
\psfrag{y14115}[cr][cr]{\PFGstyle {\small $1.4115$}}
\psfrag{y1412}[cr][cr]{\PFGstyle {\small $1.412$}}
\psfrag{y14125}[cr][cr]{\PFGstyle {\small $1.4125$}}
\psfrag{y1413}[cr][cr]{\PFGstyle {\small $1.413$}}
\psfrag{y14135}[cr][cr]{\PFGstyle {\small $1.4135$}}
\psfrag{y1414}[cr][cr]{\PFGstyle {\small $1.414$}}
\psfrag{x05}[tc][tc]{\PFGstyle $0.5$}
\psfrag{x051}[tc][tc]{\PFGstyle $0.51$}
\psfrag{x052}[tc][tc]{\PFGstyle $0.52$}
\psfrag{x053}[tc][tc]{\PFGstyle $0.53$}
\psfrag{x054}[tc][tc]{\PFGstyle $0.54$}
\psfrag{x055}[tc][tc]{\PFGstyle $0.55$}
\psfrag{x056}[tc][tc]{\PFGstyle $0.56$}
\psfrag{x057}[tc][tc]{\PFGstyle $0.57$}
\psfrag{x058}[tc][tc]{\PFGstyle $0.58$}
\psfrag{x059}[tc][tc]{\PFGstyle $0.59$}
\psfrag{x06}[tc][tc]{\PFGstyle $0.6$}
\psfrag{x061}[tc][tc]{\PFGstyle $0.61$}
\psfrag{y1}[cr][cr]{\PFGstyle $1$}
\psfrag{y11}[cr][cr]{\PFGstyle $1.1$}
\psfrag{y12}[cr][cr]{\PFGstyle $1.2$}
\psfrag{y13}[cr][cr]{\PFGstyle $1.3$}
\psfrag{y14}[cr][cr]{\PFGstyle $1.4$}
\psfrag{y15}[cr][cr]{\PFGstyle $1.5$}
\psfrag{y16}[cr][cr]{\PFGstyle $1.6$}
\psfrag{y17}[cr][cr]{\PFGstyle $1.7$}
\includegraphics[width=0.9\textwidth]{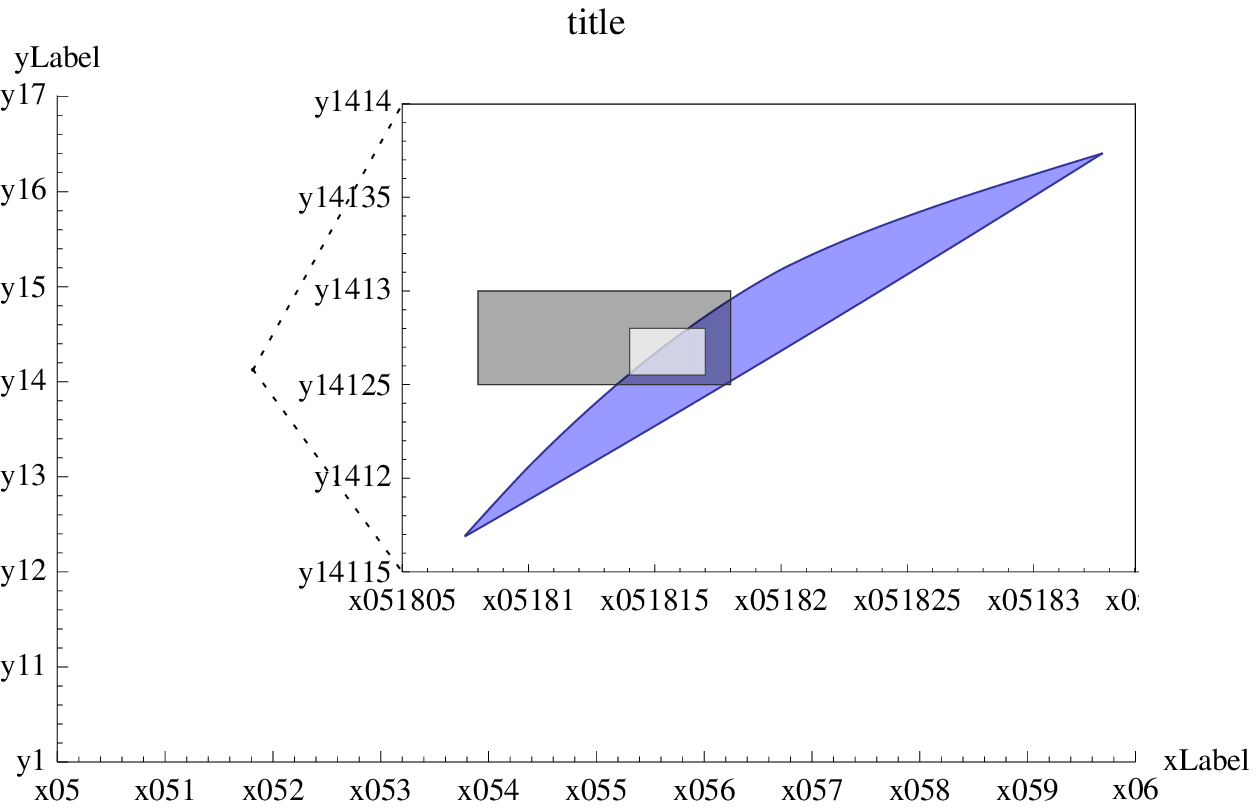}
\end{psfrags}
\caption{Allowed values of $(\De_\s,\De_\e)$ in a $\Z_2$-symmetric CFT${}_3$ containing only two relevant scalars.  The  blue region is our rigorous bound from figure~\ref{fig:allowedRegionNmax10}, computed at $n_\mathrm{max}=10,14$.  The dark grey rectangle is the Monte-Carlo prediction of \cite{Hasenbusch:2011yya}.  The light grey rectangle is the prediction of the $c$-minimization conjecture \cite{El-Showk:2014dwa}, using single-correlator results at $n_\mathrm{max}=21$.  There may be additional disconnected regions for $\De_\s\gtrsim 0.54$, as in figure~\ref{fig:nmax6MulticorrelatorRegionPlot}, but we have not computed them here.}
\label{fig:allowedRegionNmax10Summary}
\end{center}
\end{figure}

The allowed region around the Ising point shrinks further when we increase the value of $n_\text{max}$. Finding the allowed region at $n_\text{max} = 10$ ($N = 275$) is computationally intensive, so we tested only the grid of $700$ points shown in figure~\ref{fig:allowedRegionNmax10}. The disallowed points in the figure were excluded by assuming both $\De_{\s'} \ge 3$ and $\De_{\e'} \ge 3$.  On the same plot, we also show the $n_\mathrm{max}=14$ single-correlator bound on $\De_\e$ computed in \cite{El-Showk:2014dwa} using a very different optimization algorithm.  The final allowed region is the intersection of the region below the $n_\mathrm{max}=14$ curve and the region indicated by our allowed multiple correlator points.

Since the point corresponding to the 3D Ising model must lie somewhere in the allowed region, we can think of the allowed region as a rigorous prediction of the Ising model dimensions, giving $\De_\s=1/2+\eta/2=0.51820(14)$ and $\De_\e=3-1/\nu=1.4127(11)$. In figure~\ref{fig:allowedRegionNmax10Summary} we compare our rigorous bound with the best-to-date predictions using Monte Carlo simulations~\cite{Hasenbusch:2011yya} and the $c$-minimization conjecture \cite{El-Showk:2014dwa}. Although our result has uncertainties greater than $c$-minimization by a factor of ${\sim}10$ and Monte-Carlo determinations by a factor of ${\sim}3$, they still determine $\De_\s$ and $\De_\e$ with $0.03\%$ and $0.08\%$ relative uncertainty, respectively. Increasing $n_{\text{max}}$ further could potentially lead to even better determinations of $\De_\s$ and $\De_\e$. Indeed, the single correlator bound at $n_\text{max}=14$ passing through the allowed region in figure~\ref{fig:allowedRegionNmax10} indicates that the $n_\text{max}=10$ allowed region is not yet optimal. At this point, it is not even clear whether continually increasing  $n_{\text{max}}$ might lead to a finite allowed region or a single isolated point.

 We note that in our determinations we did not assume the $c$-minimization conjecture or anything similar. The only assumption besides unitarity and conformal symmetry was the existence of a $\Z_2$ symmetry and the assumption that $\s$ and $\e$ are the only relevant scalars.  It is therefore encouraging that the two methods are in such good agreement.

\section{Discussion}
\label{sec:discussion}

In this work we have elucidated the power of mixed correlators in the context of the 3D conformal bootstrap. While the simplest upper bound on the leading $\mathbb{Z}_2$-even operator dimension $\Delta_{\epsilon}$ does not differ from the single correlator bootstrap, mild assumptions about the number of relevant operators give rise to very tight constraints on the allowed values of $\Delta_{\sigma}$ and $\Delta_{\epsilon}$, almost uniquely determining their values.  Our results support the conjecture that the 3D Ising CFT is the {\it only} $\Z_2$-symmetric CFT in 3 dimensions with exactly two relevant operators.  No other such CFT has been found experimentally, and it appears that using bootstrap techniques a numerical ``proof" may be forthcoming.

Moreover by considering the mixed correlator bootstrap we are also able to gain information about the $\mathbb{Z}_2$-odd spectrum, finding a general upper bound on $\Delta_{\sigma'}$. We fully anticipate that further studies of the mixed correlator bootstrap will yield an accurate picture of the complete low-lying spectrum of the 3D Ising CFT.

There are several directions for future research. First, given the vital role that semidefinite constraints play in general formulations of the conformal bootstrap, it is important to find and implement improved algorithms for high-precision solutions of semidefinite programs of the type encountered in this work. Such improvements will make it much easier to perform broad explorations of the space of conformal field theories in general dimensions.

More concretely, it would be interesting to perform similar studies of the simplest multiple correlator constraints in $D \neq 3$, as well as in CFTs with different global symmetry groups. For example, in 2D one could understand what assumptions are needed in order to isolate the minimal model solutions, in 3D one could perform similar studies of the $O(N)$ vector models, and in 4D one could try to better understand the space of CFTs with a small number of relevant operators which may have phenomenological interest. Moreover, the time is ripe to begin including constraints from 4-point functions of operators with spin --- such studies will likely use techniques similar to what we have developed in this work.

\footnotetext{The computed points in figure~\ref{fig:allowedRegionNmax10} lie on a grid, where each row has constant $\De_\e-\De_\s$, because $\De_\e-\De_\s$ is the quantity entering the conformal blocks $g^{\De_{12},\De_{34}}_{\De,\ell}(u,v)$. Restricting it to a small number of values means we have fewer tables of blocks to compute.  We thank Slava Rychkov for this idea.}

It is also interesting to study mixed correlators in theories with supersymmetry. In particular, previous numerical bootstrap studies have focused on 4-point functions containing the lowest component of a given supersymmetry multiplet, while mixed correlators could allow one to incorporate the full constraints of supersymmetry on the external operators.\footnote{One can think about this in two ways: in components, we have four-point functions of different operators in the same SUSY multiplet; in manifestly supersymmetric notation, multiple superconformally covariant structures can appear in a three-point function.} Such studies may help to clarify the origin of the ``kink'' observed in previous studies of the 4D $\cN=1$ superconformal bootstrap~\cite{Poland:2011ey} and may also reveal rich new structure in theories with $\cN=2,4$ supersymmetry, extending the results of~\cite{Beem:2013qxa,Alday:2013opa,Alday:2014qfa}. Finally, there is significant room for incorporating mixed correlators into general analytical studies of the bootstrap, both in the context of large $N$ theories~\cite{Heemskerk:2009pn,Heemskerk:2010ty} and in constraining the spectrum at large spin~\cite{Fitzpatrick:2012yx,Komargodski:2012ek}.  

While the conformal bootstrap involving identical external operators has already shown itself to be surprisingly constraining, our results demonstrate that the larger system of mixed constraints, combined with mild assumptions about gaps, may be sufficiently powerful to uniquely locate isolated CFTs. Indeed, if one previously did not know about the 3D Ising CFT, one would have discovered it following the general logic of this paper! There may be many more isolated CFTs waiting to be discovered, perhaps theories without Lagrangian descriptions or supersymmetry. The space of such theories can be mapped out in a systematic way using the conformal bootstrap, by inputting gaps and searching for small closed allowed regions in the space of operator dimensions. There is much exploration to be done!

\clearpage
\section*{Acknowledgements}
We thank Alessandro Vichi for important discussions at the initial stages of this project. We also thank Chris Beem, Alex Dahlen, Sheer El-Showk, George Fleming, Liam Fitzpatrick, Christoph Hanselka, Joanna Huey, Jared Kaplan, Zuhair Khandker, Daliang Li, Miguel Paulos, Jo\~ao Penedones, Slava Rychkov, Leonardo Rastelli, Balt van Rees, Markus Schweighofer, and Andreas Stergiou for discussions. We are grateful to Slava Rychkov for comments on the draft.  We additionally thank the other organizers and participants in the Back to the Bootstrap 3 conference at CERN, and the organizers and participants of the New Nonperturbative Methods in QFT workshop at the KITP in Santa Barbara. The work of DSD is supported by DOE grant number DE-SC0009988.  The work of DP and FK is supported by NSF grant 1350180. The computations in this paper were run on the Bulldog computing clusters supported by the facilities and staff of the Yale University Faculty of Arts and Sciences High Performance Computing Center, as well as the Hyperion computing cluster supported by the School of Natural Sciences Computing Staff at the Institute for Advanced Study.

\clearpage
\appendix

\section{Multiple $\SU(n)$ Three-Point Structures}
\label{app:threeptglobalexample}

In this appendix, we give a concrete example of multiple structures appearing in a three-point function of operators charged under a global symmetry.
Let $G=\SU(n)$ and let $\br$ be the largest irreducible representation in $\Sym^2\mathrm{Ad}_G$, of dimension $\frac{1}{4}n^2(n-1)(n+3)$.  The symmetric tensor square of $\br$ decomposes into irreducibles as
\be
\Sym^2 \br = 2\,\br\oplus\dots.
\ee
Consequently, there are two independent three-point structures (and hence OPE coefficients) in the three-point function $\<\f_\br\f_\br\cO_{\br}\>$ when $\cO_{\br}$ has even spin.  Let us write these structures explicitly in the case where $\cO=\f$.

In terms of $\SU(n)$-indices, $\f_{ij}^{kl}$ has two symmetric upper and two symmetric lower indices, and satisfies the tracelessness condition $\f_{ij}^{il}=0$.  To write correlators of $\f$, it is convenient to use index-free notation, where we contract $\f$ with auxiliary bosonic vectors $U,\bar V$ in the fundamental and dual representations, respectively,
\be
\f(U,\bar V,x) &\equiv& \bar V^i \bar V^j U_k U_l\f_{ij}^{kl}(x).
\ee
The operator with explicit indices can be recovered by differentiating with respect to $U,\bar V$ and subtracting traces of the form $\de^k_i,\de^l_j,\de^k_j,\de^l_i$,
\be
\label{eq:recoverindices}
\f^{kl}_{ij}(x) &=& \pdr{}{U_k}\pdr{}{U_l}\pdr{}{\bar V^i}\pdr{}{\bar V^j} \f(U,\bar V,x)-\textrm{traces}.
\ee
Any expression of the form $\bar V\.U$ does not contribute after subtracting traces.  Hence, $\f(U,\bar V)$ should only be defined modulo the ideal of functions proportional to $\bar V\.U$.  Quotienting by this ideal is equivalent to restricting $\f(U,\bar V)$ to the locus $\bar V\.U=0$, so we will henceforth impose this condition.  To apply (\ref{eq:recoverindices}), we can choose an arbitrary extension of $\f(U,\bar V)$ away from $\bar V\.U=0$ and then differentiate.  Similar index-free techniques were used for classifying correlators of operators with spin in \cite{Giombi:2011rz,Costa:2011mg,SimmonsDuffin:2012uy}.

A correlator of $\f(U_m,\bar V_m,x_m)$'s must be a function of the $\SU(n)$-invariants $Q_{\bar m n}\equiv\bar V_{m}\. U_n$ (with $n\neq m$ since $\bar V_m\.U_m=0$) which is quadratic in each of the $\bar V_m, U_m$.  For a three-point function, there are two such structures consistent with permutation symmetry,
\be
\<\f(U_1,\bar V_1,x_1)\f(U_2,\bar V_2, x_2)\f(U_3,\bar V_3,x_3)\>&=& \l^1 \frac{Q_{1\bar 2}Q_{2\bar 3}Q_{3\bar 1}Q_{2\bar 1}Q_{3\bar 2}Q_{1\bar 3}}{(x_{12}x_{23}x_{31})^{\De_\f}}\nn\\
&&+\l^2\frac{\p{(Q_{1\bar 2}Q_{2\bar 3}Q_{3\bar 1})^2+(Q_{2\bar 1}Q_{3\bar 2}Q_{1\bar 3})^2}}{(x_{12}x_{23}x_{31})^{\De_\f}}.\nn\\
\ee
Each structure comes with its own OPE coefficient $\l^1,\l^2$.  The explicit $\SU(n)$-indices for this three-point function can be recovered by applying (\ref{eq:recoverindices}) for each operator.

\section{Implementation in \texttt{SDPA-GMP}}
\label{app:sdpa}

In this appendix, we follow the notation of the \texttt{SDPA} manual \cite{SDPAMANUAL}.
In section~\ref{sec:convex}, we expressed our semidefinite program in terms of the variables $a_{mn}^i$ and the positive semidefinite matrices $A_\ell, B_\ell, C_\ell, D_\ell$.  These are subject to linear constraints (\ref{eq:yequality}, \ref{eq:zequality}), where we equate coefficients of each power of $x$ on both sides.  The $a_{mn}^i$ are unconstrained.  We can write them in terms of positive variables by introducing a ``slack variable" $s\geq 0$, and writing $a_{mn}^i=b_{mn}^i-s$, where $b_{mn}^i\geq 0$.  All unknowns can now be grouped into one block-diagonal positive semidefinite matrix
\be
Y &=& \mathrm{diag}(b_{mn}^i,s,A_\ell,B_\ell,C_\ell,D_\ell).
\ee
Here, $m,n,i$ run over the $5 n_\mathrm{max}(n_\mathrm{max}+1)/2$ components of the functional $\a$, and $\ell$ runs over spins up to some large maximum value.  In this work, we take $\ell=0,1,\dots,25,26,49,50$.  Derivatives of the conformal blocks converge rapidly as $\ell\to\oo$, and in practice $\ell_\mathrm{max}\approx 50$ is enough to ensure appropriate positivity conditions for all $\ell$ (one can check this by plotting functionals at high $\ell$ once they are determined).

$Y$ is the matrix of unknowns in the ``dual" formulation of a semidefinite program defined in \cite{SDPAMANUAL}.  The equality conditions (\ref{eq:yequality}, \ref{eq:zequality}) can be expressed in the form
\be
\Tr(F_i Y) = c_i,
\label{eq:traceconditionsonY}
\ee
where $F_i$ are symmetric matrices with the same block-structure as $Y$, $i$ runs over spins $\ell$ and powers of $x$ entering (\ref{eq:yequality}, \ref{eq:zequality}), and $c_i=0$.  As a normalization condition, we turn the first constraint in (\ref{eq:constraintsformixedcorrelator}) into an equality
\be
\begin{pmatrix} 1 & 1\end{pmatrix}Z_0(0) \begin{pmatrix} 1 \\ 1 \end{pmatrix}  &=& 1,
\ee
which can also be written in the form (\ref{eq:traceconditionsonY}) with $c_i=1$, where only the entries in $F_i$ corresponding to $b_{mn}^i,s$ are nonzero.  Since we are only interested in determining whether a feasible solution exists for $Y$, and not in optimizing a particular function, we take the dual objective function $F_0$ to be identically zero.

Sometimes we have isolated operators in the OPE (for instance, in the Ising model, $\e$ is isolated from the remaining $\Z_2$-even scalars which have $\De\geq 3$).  Demanding that $\a$ be positive on the contribution of these operators gives additional semidefiniteness constraints on the variables $b_{mn}^i, s$.  To accommodate these, we extend the matrix $Y$ with semidefinite matrices $E_k$,
\be
Y = \mathrm{diag}(b_{mn}^i,s,A_\ell,B_\ell,C_\ell,D_\ell,E_k).
\ee
The semidefiniteness constraints now become equalities relating $b_{mn}^i,s$, and $E_k$, which can again be written in the form~(\ref{eq:traceconditionsonY}).  This suffices to write our semidefinite program in the form required by \texttt{SDPA-GMP}.

If our semidefinite program is feasible (i.e., if a functional $\a$ exists satisfying our constraints), then it should be possible to reduce the primal objective function to zero (the dual objective function is identically zero).  By decreasing the parameter \texttt{epsilonDash}, we can force \texttt{SDPA-GMP} to make the primal objective smaller and smaller.  Failure to decrease the primal objective below a given finite value means the SDP is infeasible.  In practice, we set \texttt{epsilonDash} very small ($\sim 10^{-30}$), and use the final value of the primal objective as a measure of whether the problem is feasible or infeasible.  Our \texttt{SDPA-GMP} parameters are summarized in table~\ref{tab:SDPAparams}.  Our condition for feasibility is $|\mathtt{primalObjective}|\leq 10^{-13}$.

\begin{table}
\begin{center}
\begin{tabular}{l | l}
parameter & value \\
\hline
\texttt{maxIteration} & $1000$ \\
\texttt{epsilonStar} & $10^{-20}$ \\
\texttt{lambdaStar} & $10^{20}$ \\
\texttt{omegaStar} & $10^{20}$ \\
\texttt{lowerBound} & $-10^{40}$ \\
\texttt{upperBound} & $10^{40}$ \\
\texttt{betaStar} & $0.1$ \\
\texttt{betaBar} & $0.3$ \\
\texttt{gammaStar} & $0.7$ \\
\texttt{epsilonDash} & $10^{-30}$ ($10^{-35}$) \\
\texttt{precision} & $300$ ($425$)
\end{tabular}
\end{center}
\caption{\texttt{SDPA-GMP} parameters used in the calculation of the operator dimension bounds. In parentheses are the values of the parameters used in the $n_\mathrm{max}=10$ computation in figure~\ref{fig:allowedRegionNmax10}.}
\label{tab:SDPAparams}
\end{table}

Our computational setup is as follows.  A \texttt{Mathematica} program computes tables of derivatives of conformal blocks using the recursion relation described in section~\ref{sec:rational}.  A separate \texttt{Mathematica} program reads these tables and writes a semidefinite program to a file in sparse \texttt{SDPA} format.   This file is read and solved by \texttt{SDPA-GMP} itself.  We have modified \texttt{SDPA-GMP} to allow checkpointing: it periodically saves its state to a file so it can be started and stopped at will.\footnote{The checkpointed version of \texttt{SDPA-GMP} is available at \\ \url{https://bitbucket.org/davidsd/sdpa-gmp-checkpointed/overview}.}  Thus, computations taking several days can be interrupted safely without having to start over again from the beginning.  The $n_\mathrm{max}=10$ computations in this work are quite time-intensive.  Writing the SDP to a file takes about 30 minutes, and solving it takes approximately 2 weeks.  The $n_\mathrm{max}=6$ computations are much less time-intensive, taking about 8 hours each.  It is extremely useful to run several computations in parallel on a computing cluster.  Checkpointing allows us to set small time-limits on each individual process, continually freeing up the cluster for jobs from other users.\footnote{Our cluster management software uses \texttt{Cloud Haskell} \cite{Epstein:2011:THC:2096148.2034690,DISTRIBUTEDPROCESS} and \texttt{MongoDB} \cite{MONGODB}.}

\clearpage
\bibliography{Biblio}{}
\bibliographystyle{utphys}

\end{document}